\documentclass[a4paper,11pt]{article}
\usepackage{amsfonts}
\usepackage{amsmath}
\usepackage{amssymb}
\usepackage{graphicx}
\usepackage{array}
\usepackage{booktabs}
\usepackage{multirow}
\usepackage{makecell}
\usepackage{float}
\pdfoutput=1 % if your are submitting a pdflatex (i.e. if you have
\usepackage{mathtools}
\usepackage{subcaption}
\usepackage{mymacros}
\usepackage{chngcntr}
\usepackage{verbatim}
\usepackage{amsthm}
\usepackage{graphics}
\usepackage{xcolor}
\usepackage{mathrsfs}
\usepackage{bbold}
\usepackage{cases}
\usepackage{epsfig}
\usepackage{epstopdf}
\usepackage{hyperref}
\usepackage{amsfonts}
\usepackage{hyperref}
\usepackage{jheppub} % for details on the use of the package, please
\usepackage[T1]{fontenc} % if needed

\hypersetup{
	colorlinks=true,
	linkcolor=blue,
	filecolor=blue,
	urlcolor=blue,
	citecolor=blue,
}

\title{\boldmath Horizon--orbit scale competition underlying chaos bound violation for spinning particles}

\author{Deyou Chen$^a$\footnote{E-mail: deyouchen@hotmail.com},}
\author{Lifang Li$^b$\footnote{E-mail: lilifang@imech.ac.cn, corresponding author.},}
\author{Chuang Yang$^a$\footnote{E-mail: chuangyangyc@hotmail.com},}
\author{Kangqiao Liu$^a$\footnote{E-mail: kqliu@xhu.edu.cn}}

\affiliation{$^a$School of Science, Xihua University, Chengdu 610039, China}
\affiliation{$^b$Center for Gravitational Wave Experiment, National Microgravity Laboratory, Institute of Mechanics, Chinese Academy of Sciences, Beijing 100190, China}

\abstract{In this paper, we investigate the chaos bound through the local radial instability of charged spinning test particles on unstable circular orbits in the black-bounce--Kerr--Newman spacetime. Our results show that violation of the bound is governed by the competition between the local orbital-instability scale and the surface gravity scale, with variations of the black hole background and the particle dynamics providing two complementary routes to the same threshold crossing. Background variations affect both scales and can drive the system into the bound-violating regime when the surface gravity is suppressed more strongly than the orbital instability. At fixed background, by contrast, the surface gravity remains unchanged, and the transition across the threshold is driven by changes in the unstable orbit dynamics induced by the particle parameters. The background-controlled and probe-dynamics-controlled routes therefore represent two realizations of a unified horizon--orbit scale competition rather than independent mechanisms.}

\begin{document} 
	\maketitle
	\flushbottom

\section{Introduction}
\label{sec1}

Unstable orbits in the strong gravity region of black holes provide a natural setting for probing the interplay between local orbital instability and horizon properties. The local radial instability of such an orbit can be characterized by the Lyapunov exponent (LE), whereas the surface gravity defines a characteristic scale associated with the event horizon. Comparing these two quantities leads to the chaos bound condition \cite{HT}
\begin{equation}
\lambda \leq \kappa ,
\label{eq1.1}
\end{equation}
\noindent where $\lambda$ denotes the LE and $\kappa$ is the black hole surface gravity. This relation is closely connected to the Maldacena--Shenker--Stanford (MSS) bound on chaos \cite{MSS}, $\lambda_L\leq 2\pi T$, which constrains the Lyapunov growth rate in thermal quantum systems and is commonly characterized through the early-time behavior of out-of-time-order correlators (OTOCs) as a diagnostic of information scrambling and quantum chaos \cite{SS,RS,MS,SSM,RSS,SSS}. In the black hole context, the surface gravity provides the corresponding horizon scale against which the classical orbital LE can be compared. The chaos bound therefore relates two quantities of different physical origin: one associated with the local instability of an orbit and the other determined by the near-horizon geometry.

Following the proposal of this bound, Zhao \textit{et al.} systematically investigated the static equilibrium of charged particles in a broad class of charged black hole spacetimes and demonstrated that the inequality can be violated under suitable conditions \cite{ZLL}. The analysis was subsequently extended by incorporating particle angular momentum \cite{LG1,LG2}, while chaotic motions in rotating black hole backgrounds were further examined in Ref.~\cite{KG1}. Subsequent studies generalized this framework to a wider range of gravitational and matter backgrounds and investigated the roles of black hole rotation \cite{KG2,KG3,KG5}, the cosmological constant \cite{KG4,GCYW2}, modified gravity \cite{FAAC1,FAAC2}, and spin--curvature coupling \cite{YCL,LCL1}. Related analyses further considered the black brane configurations \cite{LG4,DPS} and a variety of more general black hole geometries and dynamical settings \cite{DMM,LTW1,LTW2,DG,TBAZ,TBZ,TM,HMTW,HS,GT,GCYW1,LG3,JLLL,LMX,SPN,RP}. Collectively, these studies demonstrate that chaos bound behavior is sensitive to both the spacetime background and the dynamics of the probe.

Despite extensive evidence for chaos bound violation, its underlying physical origin remains unclear. Since the bound compares the orbital-instability scale with the surface gravity scale, its violation does not necessarily imply an absolute enhancement of the local orbital instability. A variation of the black hole background can modify the unstable orbit and the horizon geometry simultaneously and may suppress both the LE and the surface gravity at different rates. In particular, the bound can be violated when the surface gravity is reduced more strongly than the LE, even if the local orbital instability itself becomes weaker. A recent study identified a near-extremal geometric mechanism for the violation, in which the relative limiting behavior of unstable circular orbits and the degenerate horizon determines whether the orbital-instability scale remains finite or is suppressed together with the vanishing surface gravity \cite{TBZ}. This motivates a broader examination of how the balance between these two scales is controlled away from this limiting regime. For a spinning test particle, the spin--curvature coupling introduces nongeodesic effects into the motion and, within the pole--dipole approximation, is described by the Mathisson--Papapetrou--Dixon (MPD) equations \cite{HH}, while the charge introduces an additional electromagnetic interaction with the background field. When the black hole background is held fixed, these variations in the particle dynamics can modify the local orbital instability while leaving the surface gravity unchanged. This naturally suggests examining two complementary possibilities: whether the balance between the orbital-instability and surface gravity scales is altered by changes in the black hole background, which affect both scales, or by changes in the probe dynamics at fixed background, which act directly on the local orbital instability. A systematic comparison of these two types of variations is therefore important for clarifying how their different effects on the two competing scales can lead to chaos-bound violation. 

Black-bounce geometries provide a natural setting for examining these effects within a common two-scale picture. By introducing a nonzero regularization scale, these spacetimes replace the central singularity of conventional black hole geometries with a regular minimal-area surface and can interpolate among regular black holes, critical configurations, and horizonless traversable wormholes as the model parameters are varied \cite{SV}. The black-bounce--Kerr--Newman spacetime was constructed as a regularized extension of the Kerr--Newman geometry that interpolates between regular black hole and traversable-wormhole configurations \cite{FLMSV}. Its optical \cite{Ghosh} and scattering properties \cite{Li} have been studied. These studies have established the black-bounce--Kerr--Newman geometry as a useful framework for probing the observational and dynamical consequences of regularization in a rotating and charged spacetime. The regularization deformation modifies both the horizon structure and the unstable-orbit dynamics, thereby affecting the orbital-instability and surface gravity scales entering the chaos bound. With rotation and charge included, this spacetime therefore provides a useful setting for comparing background variations, which generally modify both scales, with probe-dynamics variations at fixed background, which affect the local orbital instability while leaving the surface gravity unchanged.

In this work, we investigate the local radial instability of charged spinning test particles on unstable circular orbits in the black-bounce--Kerr--Newman spacetime and examine the chaos bound by comparing the LE with the black hole surface gravity. The particle motion is restricted to the equatorial plane, with its spin taken to be parallel or antiparallel to the symmetry axis. After determining the unstable circular orbits from the radial effective potential, we analyze how the black hole and particle parameters affect the LE and the onset of chaos bound violation. Our analysis is organized to compare two complementary classes of variations within the same competition between the orbital-instability and surface gravity scales. Variations of the black hole background generally modify both scales, whereas variations of the particle total angular momentum, spin, and charge at fixed background leave the surface gravity unchanged and thereby isolate their effects on the unstable-orbit dynamics. We further compare the two orbital branches of the rotating spacetime and examine how the spin--curvature coupling modifies the local orbital instability and shifts the critical threshold for the violation in a branch-dependent manner.

The remainder of this paper is organized as follows. In Sec.~\ref{sec2}, we introduce the black-bounce--Kerr--Newman geometry and derive the equations of motion for a charged spinning test particle in this spacetime. We then construct the radial effective potential and obtain the corresponding LE. In Sec.~\ref{sec3}, we examine how the black hole background and particle parameters affect chaos bound violation and compare their respective effects on the orbital-instability and surface gravity scales. Section~\ref{sec4} summarizes our main results and conclusions.

\section{Dynamics of spinning particles in the black-bounce--Kerr--Newman spacetime}
\label{sec2}

\subsection{Black-bounce--Kerr--Newman geometry}
\label{sec2.1}

The black-bounce-Kerr--Newman geometry introduced in Ref.~\cite{FLMSV} is obtained by replacing the Kerr--Newman radial function with its regularized counterpart, $r\rightarrow\sqrt{r^2+\ell^2}$. In Boyer--Lindquist-like coordinates, the metric can be written as

\begin{equation} 
\begin{split} ds^2={}&-\frac{\Delta}{\rho^2}\left(a\sin^2\theta\,d\phi-dt\right)^2+\frac{\sin^2\theta}{\rho^2} \left[\left(r^2+\ell^2+a^2\right)d\phi-a\,dt\right]^2 \\ 
&+\frac{\rho^2}{\Delta}\,dr^2 +\rho^2\,d\theta^2 , 
\end{split} \label{metric1} 
\end{equation}

\noindent together with the electromagnetic four-potential

\begin{equation} 
A_{\mu} =-\frac{Q\sqrt{r^2+\ell^2}}{\rho^2}\left(1,0,0,-a\sin^2\theta\right), \label{ep} 
\end{equation}

\noindent where

\begin{align} 
\rho^2 &=r^2+\ell^2+a^2\cos^2\theta,\\ \Delta&=r^2+\ell^2+a^2-2M\sqrt{r^2+\ell^2}+Q^2 . \label{metric2} 
\end{align}

\noindent Here, \(M\), \(a\), and \(Q\) denote the mass, rotation parameter, and electric charge, respectively, while $\ell>0$ is the regularization length scale governing the black-bounce deformation. As in the original black-bounce construction \cite{SV}, the radial coordinate extends over the entire real line. In the rotating charged generalization \cite{FLMSV}, the surface $r=0$ provides a regular connection between the two asymptotic regions.

The Killing horizons are determined by the real roots of $\Delta=0$. In the parameter regime
$Q^{2}+a^{2}\leq M^{2}$, it is convenient to introduce

\begin{equation}
\ell_{\pm}=M\pm\sqrt{M^{2}-Q^{2}-a^{2}},
\label{ell+}
\end{equation}

\noindent where $\ell_{+}$ and $\ell_{-}$ coincide with the outer and inner horizon radii, respectively, of the Kerr--Newman spacetime with the same values of $M$, $a$, and $Q$. The horizons of the black-bounce--Kerr--Newman geometry are then located at

\begin{equation}
r_{H}^{(S_{1},\pm)} =S_{1}\sqrt{\ell_{\pm}^{2}-\ell^{2}}.
\label{rh}
\end{equation}

\noindent  Here, $S_{1}=+1$ and $S_{1}=-1$ identify the two asymptotic regions $r>0$ and $r<0$, respectively, while the labels $+$ and $-$ refer to the outer and inner horizon branches. In the $r>0$ asymptotic region, the corresponding horizon radii reduce to

\begin{equation}
r_{\pm} =\sqrt{\ell_{\pm}^{2}-\ell^{2}}.
\label{rpm}
\end{equation}

\noindent  The outer horizons exist for $\ell\leq\ell_{+}$, whereas the existence of the inner horizons further requires $\ell\leq\ell_{-}$. Since $\ell_{-}\leq\ell_{+}$, the horizon structure in the subextremal regime, $Q^{2}+a^{2}<M^{2}$, can be classified as follows~\cite{FLMSV}:

\begin{equation}
\begin{cases}
0<\ell<\ell_{-},
&
\text{both the inner and outer horizon pairs exist},
\\[1mm]
\ell_{-}<\ell<\ell_{+},
&
\text{only the outer horizon pair exists},
\\[1mm]
\ell>\ell_{+},
&
\text{the spacetime is horizonless}.
\end{cases}
\label{horizon-classification}
\end{equation}

\noindent  At the critical value $\ell=\ell_{-}$, the two inner horizons coalesce at the bounce surface $r=0$. Likewise, when $\ell=\ell_{+}$, the two outer horizons merge at $r=0$, producing a critical configuration with vanishing surface gravity that separates the black hole and wormhole sectors. For $\ell>\ell_{+}$, no Killing horizon is present, and the resulting regular geometry describes a two-way traversable wormhole.

In the extremal case, $Q^{2}+a^{2}=M^{2}$, the inner and outer horizon branches coincide. For $0<\ell<M$, the resulting degenerate horizons are located at

\begin{equation}
r_{H}=\pm\sqrt{M^{2}-\ell^{2}}.
\end{equation}

\noindent At the critical value $\ell=M$, the two degenerate horizons coalesce at the bounce surface $r=0$. For $\ell>M$, no real roots of $\Delta=0$ exist, and the geometry describes a horizonless two-way traversable wormhole. In the superextremal regime, $Q^{2}+a^{2}>M^{2}$,  the spacetime possesses no Killing horizons for any value of $\ell$. In contrast to the superextremal Kerr--Newman geometry, however, the absence of horizons does not expose a curvature singularity. For $\ell>0$, the Kerr--Newman singularity is regularized by the black-bounce deformation, and the resulting spacetime is a regular, horizonless geometry that can be interpreted as a traversable wormhole. 

For convenience in the following numerical analysis, we introduce the dimensionless deformation parameter 

\begin{equation}
\eta=\frac{\ell}{\ell_{+}},
\label{eta-definition}
\end{equation}

\noindent which measures the regularization scale relative to the critical value $\ell_{+}$ and therefore quantifies the proximity of the geometry to the black hole--wormhole transition. This parametrization is well defined for $Q^{2}+a^{2}\leq M^{2}$ at fixed $M$, $a$, and $Q$. The black hole sector corresponds to $0\leq\eta<1$, while $\eta=1$ identifies the critical configuration at which the outer horizons coalesce at the bounce surface. Configurations with $\eta>1$ lie in the horizonless wormhole sector. Thus, within the black hole regime, increasing $\eta$ drives the geometry toward the black hole--wormhole transition. Using the horizon and surface gravity expressions derived in~\cite{FLMSV}, the outer horizon radius and surface gravity can be recast in terms of $\eta$ as

\begin{align} 
r_{+} &= \ell_{+}\sqrt{1-\eta^{2}},\label{oh-eta}\\ 
\kappa &=\kappa^{\mathrm{KN}}\sqrt{1-\eta^{2}}, \label{sf}
\end{align}

\noindent where

\begin{equation}
\kappa^{\mathrm{KN}}
=
\frac{\sqrt{M^{2}-Q^{2}-a^{2}}}
{\left(M+\sqrt{M^{2}-Q^{2}-a^{2}}\right)^{2}+a^{2}}
\end{equation}

\noindent is the surface gravity of the Kerr--Newman black hole with the same values of $M$, $a$, and $Q$. The black-bounce deformation therefore reduces both the outer horizon radius and the surface gravity relative to their Kerr--Newman counterparts. In the Kerr--Newman limit, $\eta\to0$, one recovers

\begin{equation}
r_{+}\to\ell_{+},
\qquad
\kappa\to\kappa^{\mathrm{KN}}.
\end{equation}

\noindent By contrast, as the critical black hole--wormhole configuration is approached, $\eta\to1^{-}$, both quantities vanish.

\subsection{Equations of motion for a spinning test particle}
\label{sec2.2}

Within the pole--dipole approximation, the dynamics of a charged spinning test particle in curved spacetime is governed by the MPD equations \cite{HH},
 
\begin{align}
\frac{D\tilde{p}^{\mu}}{D\tau} &=-\frac{1}{2}R^{\mu}{}_{\nu\alpha\beta}u^{\nu}\tilde{S}^{\alpha\beta}
+\tilde{q}F^{\mu}{}_{\nu}u^{\nu},\label{eq2.01}\\
\frac{D\tilde{S}^{\mu\nu}}{D\tau}&=\tilde{p}^{\mu}u^{\nu}-u^{\mu}\tilde{p}^{\nu},\label{eq2.02}
\end{align}

\noindent where \(D/D\tau\) denotes the covariant derivative along the particle worldline, parametrized by \(\tau\). The quantities \(\tilde{p}^{\mu}\), \(u^{\mu}=dx^{\mu}/d\tau\), \(\tilde{S}^{\mu\nu}\), and \(\tilde{q}\) represent the four-momentum, four-velocity, antisymmetric spin tensor, and particle charge, respectively. In addition, \(R^{\mu}{}_{\nu\alpha\beta}\) is the Riemann curvature tensor, while \(F_{\mu\nu}=\partial_{\mu}A_{\nu}-\partial_{\nu}A_{\mu}\) is the electromagnetic field tensor.
We retain only the monopole electromagnetic interaction associated with the particle charge and neglect electromagnetic dipole and higher-multipole couplings.

For convenience, we introduce the momentum, spin, and charge normalized by the particle mass,
 
\begin{equation}
p^{\mu} =\frac{\tilde{p}^{\mu}}{m},\qquad S^{\mu\nu}=\frac{\tilde{S}^{\mu\nu}}{m},
\qquad q =\frac{\tilde{q}}{m},
\label{eqnorm}
\end{equation}
 
\noindent where \(m\) denotes the particle mass. Since the MPD equations alone do not uniquely determine the representative worldline of an extended spinning body, a spin supplementary condition is required to fix the centroid \cite{WT,HR,AO,KS,ZGWYL,ZW}. In this work, we adopt the Tulczyjew--Dixon spin supplementary condition (TDSSC) \cite{WT,HR}. This selects the centroid associated with the zero-momentum frame. In terms of the mass-normalized momentum and spin introduced above, the TDSSC retains the form
 
\begin{equation}
S^{\mu\nu}p_{\nu}=0.
\label{eq2.03}
\end{equation}
 
\noindent The normalized momentum and spin satisfy the invariant relations
 
\begin{align}
p^{\mu}p_{\mu}&=-1,
\label{eq2.8}
\\
\frac{1}{2}S_{\mu\nu}S^{\mu\nu}&=S^{2}.
\label{eq2.9}
\end{align}
 
\noindent Here, \(|S|\) denotes the spin magnitude per unit particle mass, while the sign of \(S\) specifies the spin orientation relative to the \(z\)-axis. We choose \(S>0\) for spin aligned with the positive \(z\)-direction and \(S<0\) for the corresponding antialigned configuration. This convention allows the orientation dependence of the spin--curvature coupling to be encoded in a single signed spin parameter. 

We restrict the particle motion to the equatorial plane, \(\theta=\pi/2\), and take its spin to be normal to this plane, i.e., parallel or antiparallel to the symmetry axis. This choice is consistent with the reflection symmetry of the spacetime about the equatorial plane and reduces the MPD dynamics to the symmetry-preserving equatorial sector considered throughout this work. Accordingly, the polar component of the momentum vanishes, while all spin-tensor components involving the \(\theta\) direction are set to zero,
 
\begin{equation}
p^{\theta}=0, \qquad S^{\theta\mu}=0. 
\label{eq2.10}
\end{equation}
 
\noindent 
Thus, the spin tensor is completely characterized by the three components \(S^{t\phi}\), \(S^{tr}\), and \(S^{r\phi}\). These components are not independent, since they are constrained by the TDSSC together with the normalization of the particle momentum and spin. Combining these conditions with the normalization conditions in Eqs.~\eqref{eq2.8} and \eqref{eq2.9}, the nonvanishing components of the spin tensor can be expressed directly in terms of the covariant momentum components as
 
\begin{equation}
S^{t\phi}=2\sigma p_{r}, \qquad S^{tr}=-2\sigma p_{\phi}, \qquad S^{r\phi}=-2\sigma p_{t},
\label{eq2.12}
\end{equation}
 
\noindent where
 
\begin{equation}
\sigma=
\frac{S}
{2\sqrt{g_{rr}\left(g_{t\phi}^{2}-g_{tt}g_{\phi\phi}\right)}}.
\label{eq2.11}
\end{equation}
Here \(S\) denotes the signed spin parameter of the particle, with its sign distinguishing the two possible orientations with respect to the symmetry axis. The factor \(\sigma\) incorporates both the spin magnitude and the local metric geometry, and therefore provides the geometrical normalization relating the spin-tensor components to the particle momentum. Equations~\eqref{eq2.10}--\eqref{eq2.12} consequently reduce the spin degrees of freedom to a single signed parameter \(S\), which considerably simplifies the subsequent derivation of the conserved quantities and equations of motion.

The stationarity and axisymmetry of the spacetime are generated by the Killing vector fields, $\xi_{(t)}^a=\left(\frac{\partial}{\partial t}\right)^a$ and $\xi_{(\phi)}^a=\left(\frac{\partial}{\partial \phi}\right)^a$, respectively. These symmetries give rise to the conserved energy and total angular momentum. On the equatorial plane, these conserved quantities are given by

\begin{align}
E&= -p_{t}+\sigma p_{t}g_{t\phi}^{\prime}-\sigma p_{\phi}g_{tt}^{\prime}-qA_{t},\label{eq2.13}\\
J&=p_{\phi}+\sigma p_{\phi}g_{t\phi}^{\prime}-\sigma p_{t}g_{\phi\phi}^{\prime}+qA_{\phi}, \label{eq2.14}
\end{align}

\noindent where a prime denotes differentiation with respect to \(r\). Here, \(E\) and \(J\) denote the specific energy and specific total angular momentum, respectively. The latter contains contributions from the orbital motion, the particle spin, and the electromagnetic coupling. With the standard orientation of the azimuthal coordinate, \(J>0\) and \(J<0\) label total angular momentum directed along and opposite to the positive \(z\)-axis, respectively. Solving Eqs.~\eqref{eq2.13} and \eqref{eq2.14} for the covariant momentum components \(p_t\) and \(p_\phi\), we obtain 

\begin{align}
p_{t}&= -\frac{(1+\sigma g_{t\phi}^{\prime})	\left(E+qA_{t}\right)+\sigma g_{tt}^{\prime}
	\left(J-qA_{\phi}\right)}{1-\sigma^{2}\left(\left(g_{t\phi}^{\prime}\right)^{2}-g_{tt}^{\prime}g_{\phi\phi}^{\prime}\right)},\label{eq2.17}\\
p_{\phi}&=\frac{(1-\sigma g_{t\phi}^{\prime})\left(J-qA_{\phi}\right)-\sigma g_{\phi\phi}^{\prime}\left(E+qA_{t}\right)}{1-\sigma^{2}\left(\left(g_{t\phi}^{\prime}\right)^{2}
	-g_{tt}^{\prime}g_{\phi\phi}^{\prime}\right)}.\label{eq2.18}
\end{align}
 
\noindent  The common denominator in the above equations must remain nonzero within the physically admissible region. The radial momentum follows from the mass-shell condition \(p^{\mu}p_{\mu}=-1\). On the equatorial plane, substituting Eqs.~\eqref{eq2.17} and \eqref{eq2.18} gives 

\begin{equation}
p_{r}^{2} = g_{rr}\left(-1+\frac{g_{\phi\phi}p_{t}^{2}-2g_{t\phi}p_{t}p_{\phi}+g_{tt}p_{\phi}^{2}}
{g_{t\phi}^{2}-g_{tt}g_{\phi\phi}}\right),
\label{eq2.19}
\end{equation}

\noindent  which determines the radial momentum in terms of the conserved quantities and particle parameters and provides the basis for analyzing the radial dynamics and constructing the corresponding effective potential.

For a spinning particle, its four-momentum $p^{\mu}$ is, in general, not parallel to the four-velocity $u^{\mu}$. The coordinate velocities must therefore be obtained independently from the spin-evolution equation rather than being identified directly with ratios of momentum components. Taking the coordinate time $t$ as the evolution parameter and defining $\dot{r}=dr/dt$ and $\dot{\phi}=d\phi/dt$, and considering the Eq.~\eqref{eqnorm}, Eq.~\eqref{eq2.02} gives 

\begin{eqnarray}
\frac{DS^{tr}}{D t}&=&p^t\dot r-p^r\nonumber\\
&=&\sigma\left(	R_{\phi t\alpha\beta}S^{\alpha\beta}+	R_{\phi r\alpha\beta}S^{\alpha\beta}\dot r+R_{\phi\phi\alpha\beta}S^{\alpha\beta}\dot\phi\right)+2q\sigma A'_\phi\dot r, \label{eq2.21}\\
\frac{DS^{t\phi}}{D t} 	&=&	p^t\dot\phi-p^\phi \nonumber\\ &=&-\sigma\left(R_{rt\alpha\beta}S^{\alpha\beta}+R_{rr\alpha\beta}S^{\alpha\beta}\dot r	+	R_{r\phi\alpha\beta}S^{\alpha\beta}\dot\phi\right)+ 2q\sigma\left(A'_t+A'_\phi\dot\phi\right).
\label{eq2.22}
\end{eqnarray}

\noindent In obtaining the second equalities in Eqs.~\eqref{eq2.21} and \eqref{eq2.22}, we have used the momentum-evolution equation \eqref{eq2.01} together with the spin-tensor components in Eq.~\eqref{eq2.12}. These relations form a coupled linear system for \(\dot{r}\) and \(\dot{\phi}\). Solving this system gives

\begin{eqnarray}
\dot r= \frac{b_1}{a_1}, \label{eq2.23}\\
\dot\phi=\frac{b_2}{a_2}, \label{eq2.24}
\end{eqnarray}

\noindent where

\begin{align}
a_1={}&a_2=p^t+\frac{S}{2\mathcal{R}^4}\Bigl(p^tS\mathcal{A}-a p^\phi S\mathcal{B}+2aqQr\mathcal{R}^2\Bigr),\label{a1}\\
b_1
={}&p^r-\frac{rp^rS^2} {2\mathcal{R}^3}\Bigl(2r\left(a^2-\Delta\right)+\mathcal{R}\Delta'\Bigr),\label{b1}\\
b_2={}&p^\phi-\frac{S}{2\mathcal{R}^4}\Bigl(p^\phi S\mathcal{C}-a p^t S\mathcal{D}-2qQr\mathcal{R}^2\Bigr),
\label{b2}
\end{align}
with
\begin{align}
\mathcal{R}={}&\ell^2+r^2,\label{R}\\
\mathcal{A}={}&2\Bigl(a^2\left(4r^2-\ell^2\right)+\mathcal{R}\left(r^2-\ell^2\right)\Bigr)\left(\Delta-a^2\right)-r\mathcal{R}\left(5a^2+\mathcal{R}\right)\Delta'+a^2\mathcal{R}^2\Delta'',\label{A}\\
\mathcal{B}={}&2\Bigl(a^2\left(4r^2-\ell^2\right)+2\mathcal{R}\left(2r^2-\ell^2\right)\Bigr)\left(\Delta-a^2\right)+2\ell^2\mathcal{R}^2 \nonumber\\
& - \mathcal{R}\left(a^2+\mathcal{R}\right)(5r\Delta' -\mathcal{R}\Delta''), \label{B}\\
\mathcal{C}={}&2\Bigl(a^2\left(4r^2-\ell^2\right)+\mathcal{R}\left(3r^2-\ell^2\right)\Bigr)
\left(\Delta-a^2\right) -r\mathcal{R}\left(5a^2+4\mathcal{R}\right)\Delta' \nonumber\\
& +\mathcal{R}^2\left(a^2+\mathcal{R}\right)\Delta'', \label{C}\\
\mathcal{D}={}&2\left(4r^2-\ell^2\right)\left(\Delta-a^2\right)-5r\mathcal{R}\Delta'+\mathcal{R}^2\Delta''. \label{D}
\end{align}

\noindent After imposing the TDSSC and expressing the four-velocity in terms of the momentum, the terms proportional to $qQS$ arise from the interplay between the electromagnetic interaction and the spin-dependent momentum--velocity relation, whereas the terms quadratic in $S$ reflect curvature-induced corrections associated with the spin--curvature coupling. Since the radial effective potential and the stability of circular orbits are determined by the radial dynamics, the following analysis will primarily focus on Eq.~\eqref{eq2.23}.

\subsection{Orbital instability and Lyapunov exponent}
\label{sec2.3}

The local radial stability of a circular trajectory can be characterized by the LE, which measures the rate at which a small radial displacement departs from the reference orbit \cite{CMBWZ}. Choosing the coordinate time as the evolution parameter, we rewrite the radial equation in the form
 
\begin{equation}
\frac{1}{2}m\dot{r}^{\,2} +\mathcal{V}_{\mathrm{eff}}(r)=0,
\label{eq2.3.1}
\end{equation}
 
\noindent with $\mathcal{V}_{\mathrm{eff}}(r)=-\frac{m}{2}\left(\frac{b_{1}}{a_{1}}\right)^{2}$,  where Eq.~\eqref{eq2.23} has been used. Although \(\mathcal{V}_{\mathrm{eff}}\) is constructed directly from the coordinate-time radial velocity, it provides a useful local description of the radial dynamics. For fixed black hole parameters and particle parameters \(J\) and \(S\), the conserved energy \(E\) is not prescribed independently. Instead, \(E\) and the circular orbit radius \(r_0\) are determined simultaneously for each parameter set from the conditions
 
\begin{equation}
\mathcal{V}_{\mathrm{eff}}(r_{0})=0,
\qquad
\left. \frac{\partial\mathcal{V}_{eff}\left(r\right)}{\partial r}\right|_{r=r_{0}}=0.
\label{eq2.3.3}
\end{equation}
 
\noindent The sign of the second derivative distinguishes stable and unstable configurations. In the present analysis, we focus on unstable circular orbits satisfying
 
\begin{equation}
\left.\frac{\partial^{2}\mathcal{V}_{eff}(r)}{\partial r^{2}}\right|_{r=r_0}<0.
\label{eq2.3.31}
\end{equation}

To extract the corresponding instability rate, we perturb the orbit as
\(r(t)=r_{0}+\epsilon(t)\), with \(|\epsilon|\ll r_{0}\). Expanding Eq.~\eqref{eq2.3.1} around \(r_{0}\) and retaining terms up to quadratic order gives
 
\begin{equation}
\frac{1}{2}m\dot{\epsilon}^{\,2}
+\frac{1}{2}\mathcal{V}_{\mathrm{eff}}^{\prime\prime}(r_{0})\epsilon^{2}=0.
\label{eq2.3.4}
\end{equation}
 
\noindent Taking another derivative of the above equation with respect to the coordinate time, we obtain

\begin{equation}
m\ddot{\epsilon}+\mathcal{V}_{\mathrm{eff}}^{\prime\prime}(r_{0})\epsilon=0,
\end{equation}

\noindent which can be equivalently written as $\ddot{\epsilon}-\lambda^{2}\epsilon=0$, where
 
\begin{align}
\lambda^{2}&=-\frac{1}{m}\mathcal{V}_{\mathrm{eff}}^{\prime\prime}(r_{0})=
\frac{1}{2}\left.\frac{d^{2}}{dr^{2}}\left(\frac{b_{1}}{a_{1}}\right)^{2}\right|_{r=r_{0}}
\label{le}
\end{align}
 
\noindent is the squared LE characterizing the local radial instability of the circular orbit~\cite{JLLL,CHL}. For an unstable orbit, \(\mathcal{V}_{\mathrm{eff}}^{\prime\prime}(r_{0})<0\) and hence $\lambda^{2}>0$, yielding $\epsilon(t)\propto e^{\pm\lambda t}$. The growing mode describes the radial instability, while $\lambda^{-1}$ sets its characteristic time scale.

\section{Horizon--orbit scale competition and chaos bound violation}
\label{sec3}

\begin{figure*}[htbp]
	\centering
	\subcaptionbox{$J=-6.00$}{\includegraphics[width=0.42\textwidth]{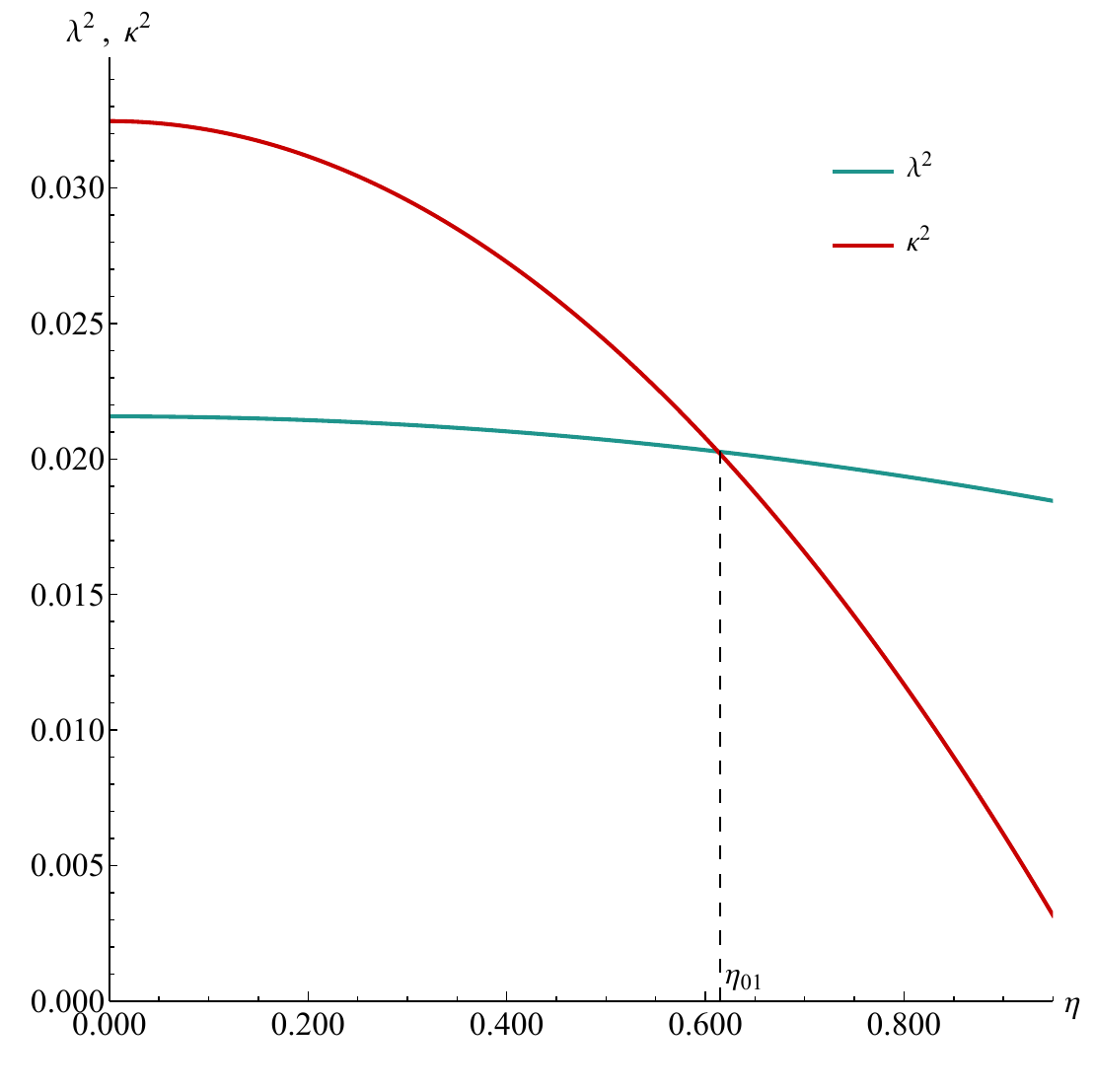}}
	\subcaptionbox{$J=6.00$}{\includegraphics[width=0.42\textwidth]{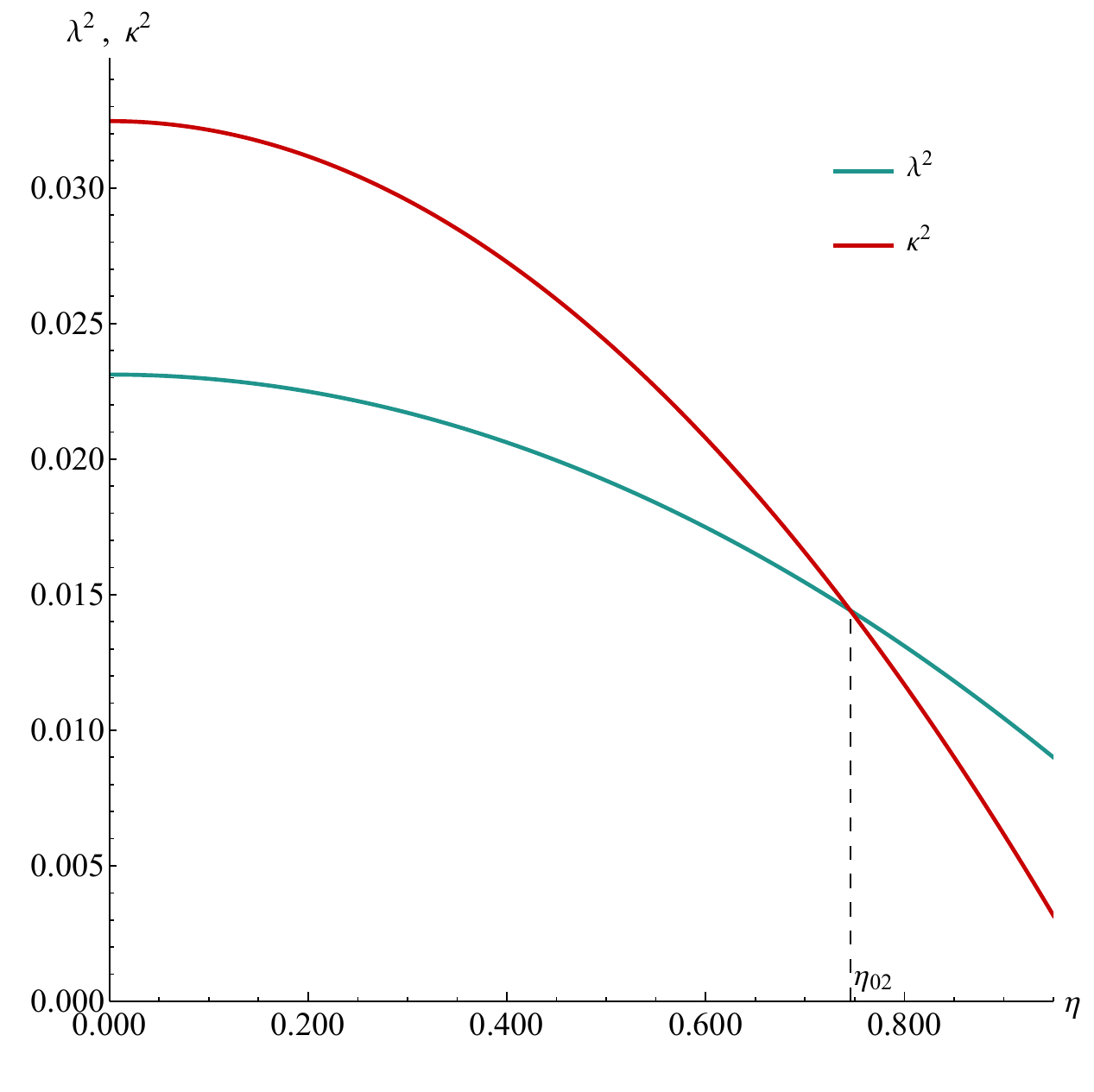}}\\[5mm]
	\subcaptionbox{$J=-6.00$}{\includegraphics[width=0.42\textwidth]{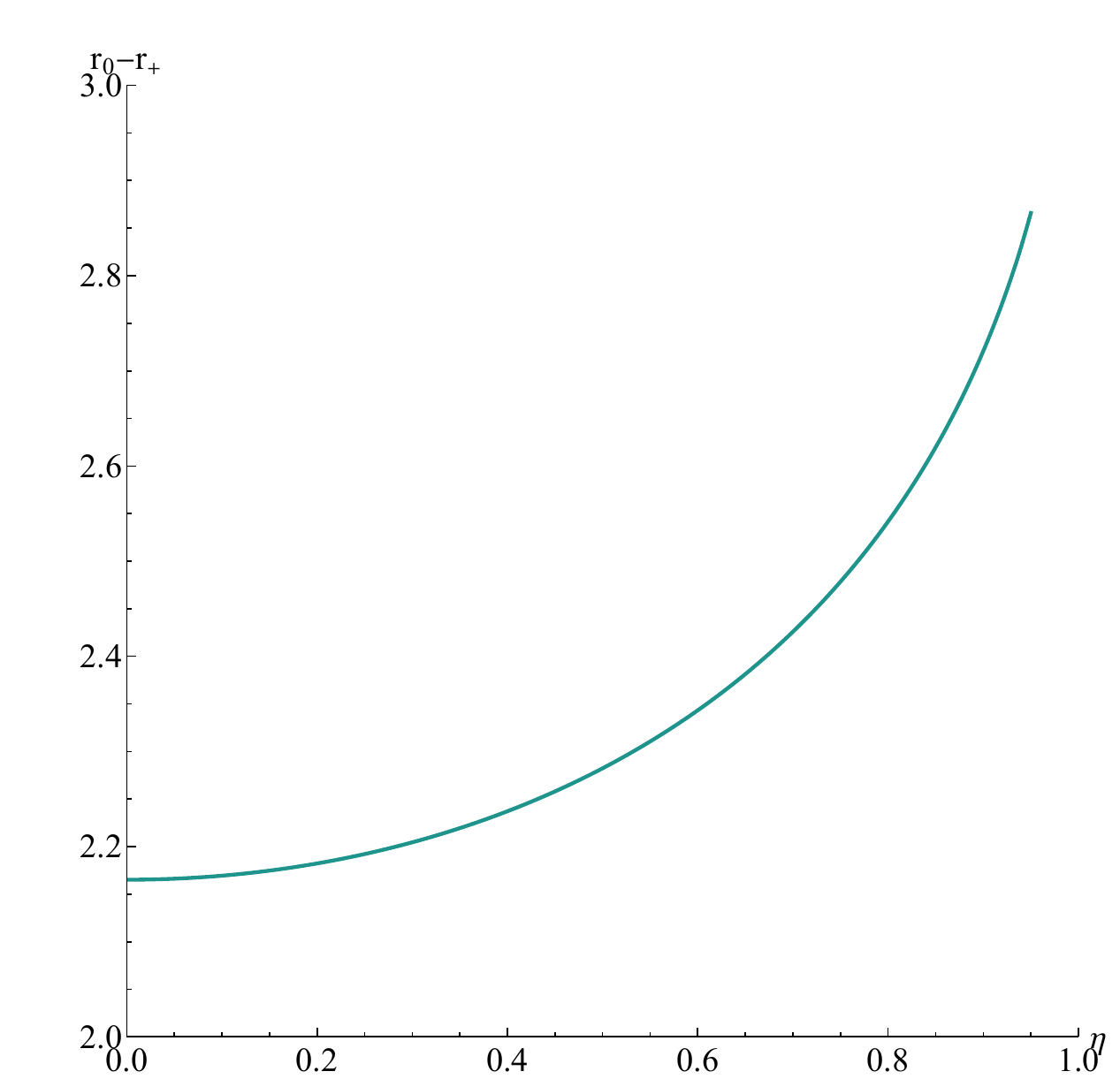}}
	\subcaptionbox{$J=6.00$}{\includegraphics[width=0.42\textwidth]{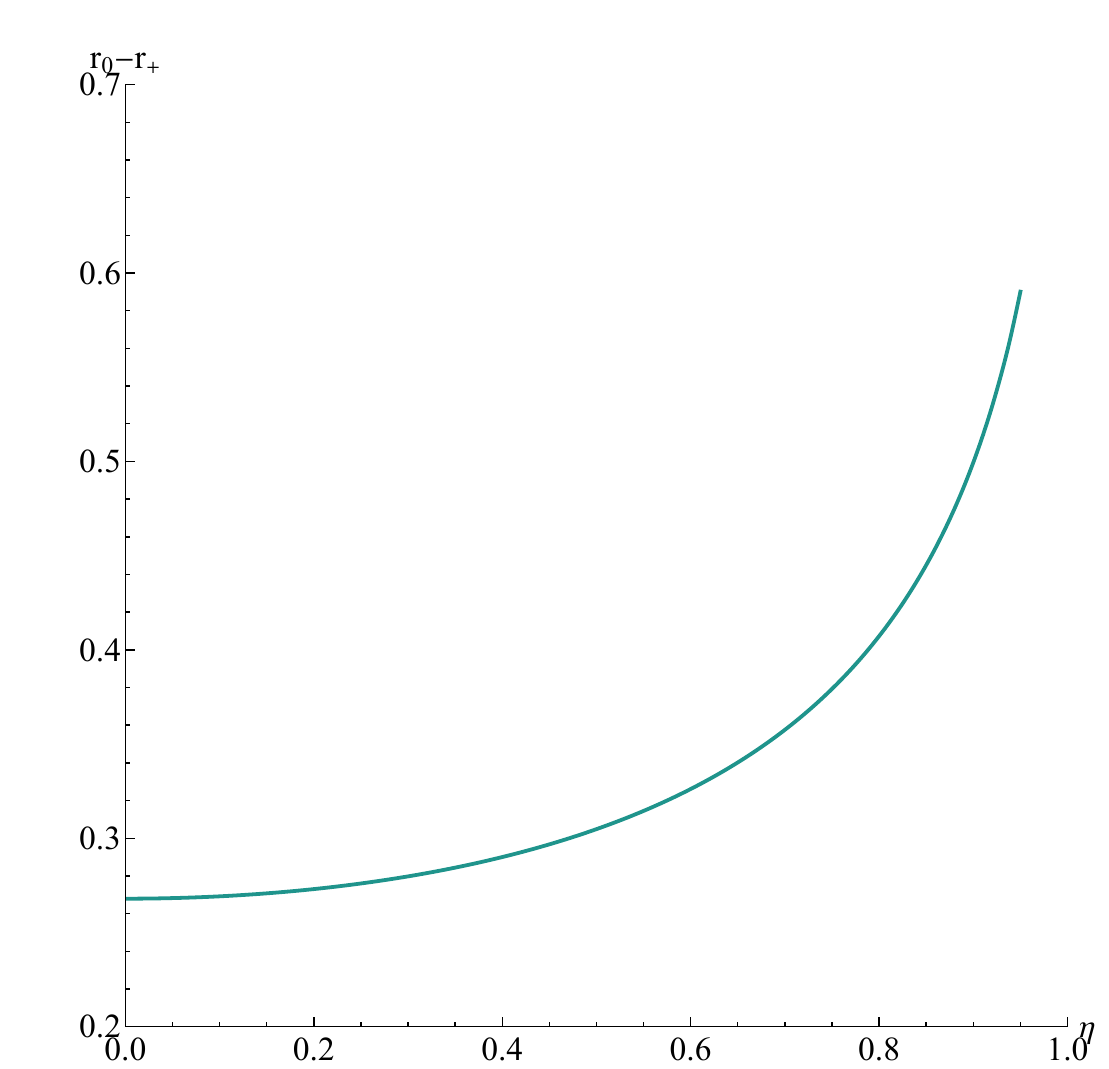}}
	\caption{Effects of the dimensionless deformation parameter \(\eta\) on the squared LE, squared surface gravity, and orbit--horizon separation for the two angular momentum branches.}
	\label{fig1}
\end{figure*}

\begin{figure*}[htbp]
	\centering
	\subcaptionbox{$J=-6.00$}{\includegraphics[width=0.42\textwidth]{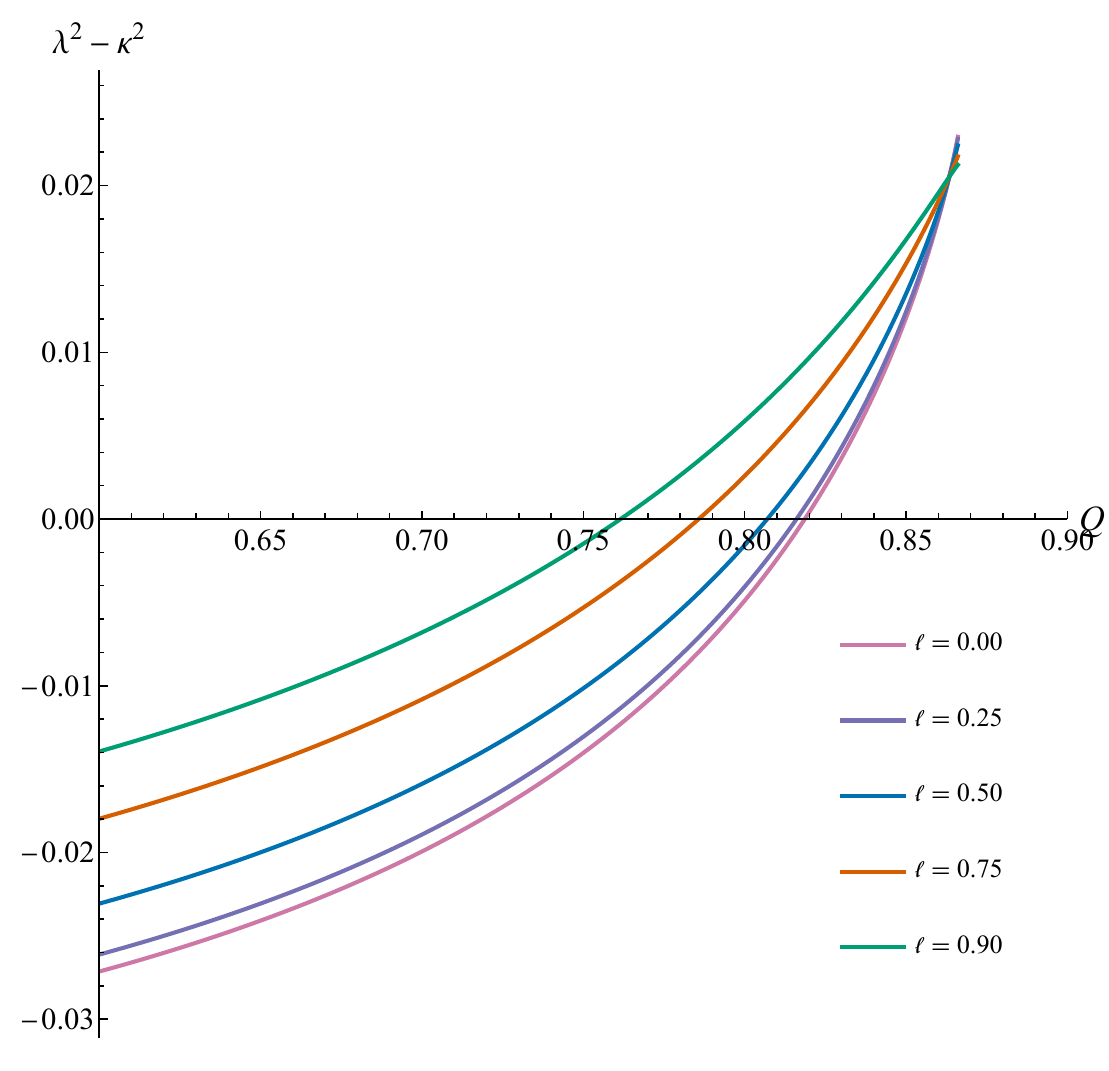}}
	\subcaptionbox{$J=6.00$}{\includegraphics[width=0.42\textwidth]{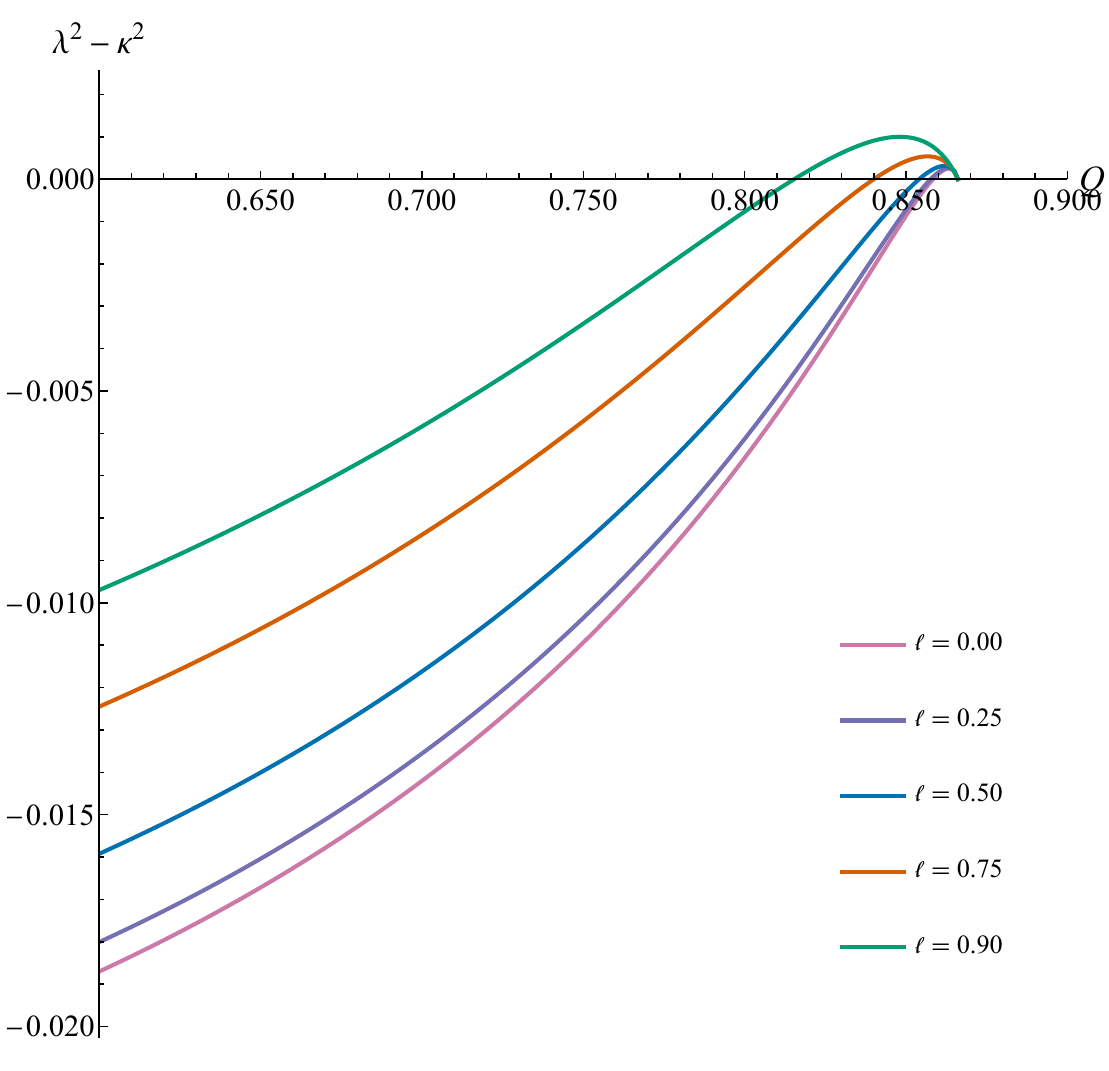}}
	\caption{Effect of the black hole charge on $\Delta_{\lambda\kappa}$.}
	\label{fig2}
\end{figure*}

\begin{figure*}[htbp]
	\centering
	\subcaptionbox{}{\includegraphics[width=0.6\textwidth]{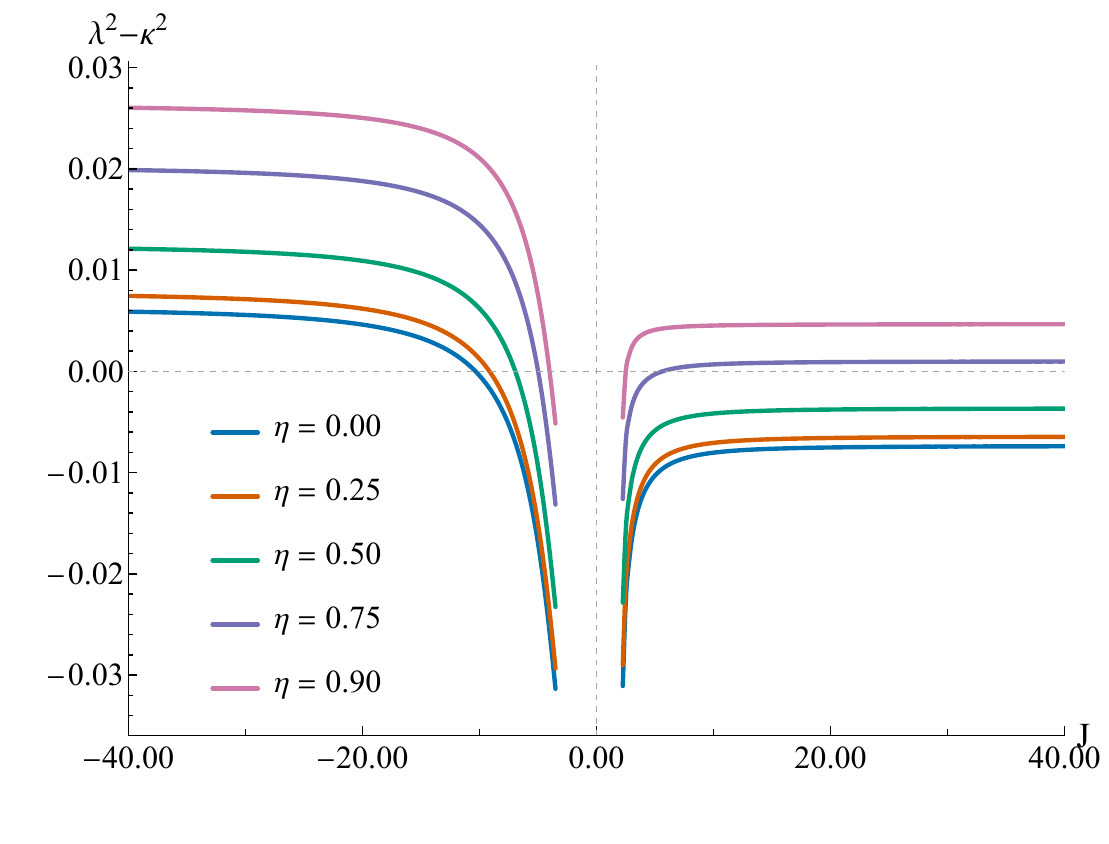}}
	\caption{Effect of the particle's total angular momentum on $\Delta_{\lambda\kappa}$. }
	\label{fig3}
\end{figure*}

\begin{figure*}[htbp]
	\centering
	\subcaptionbox{}{\includegraphics[width=0.42\textwidth]{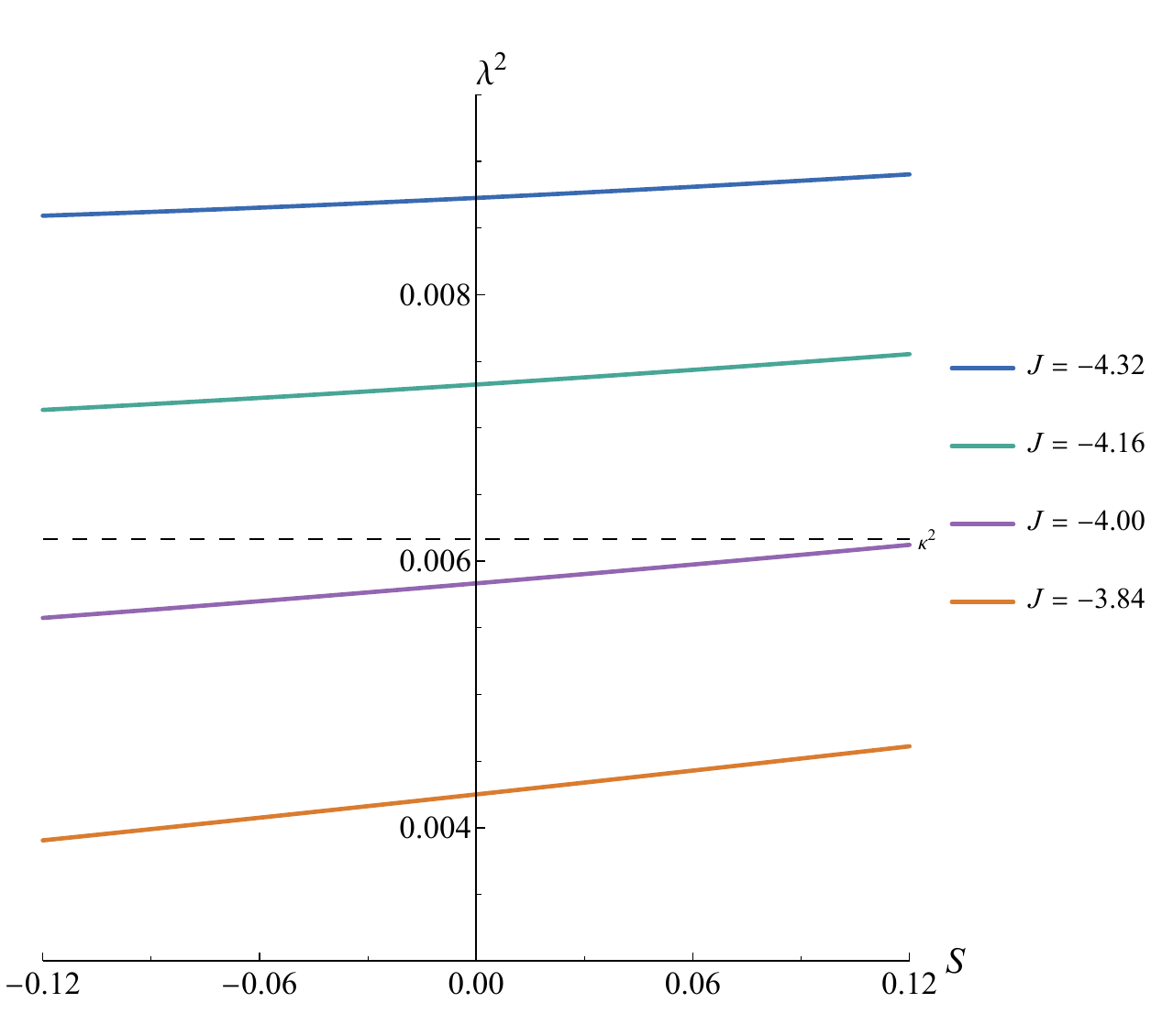}}
	\subcaptionbox{}{\includegraphics[width=0.42\textwidth]{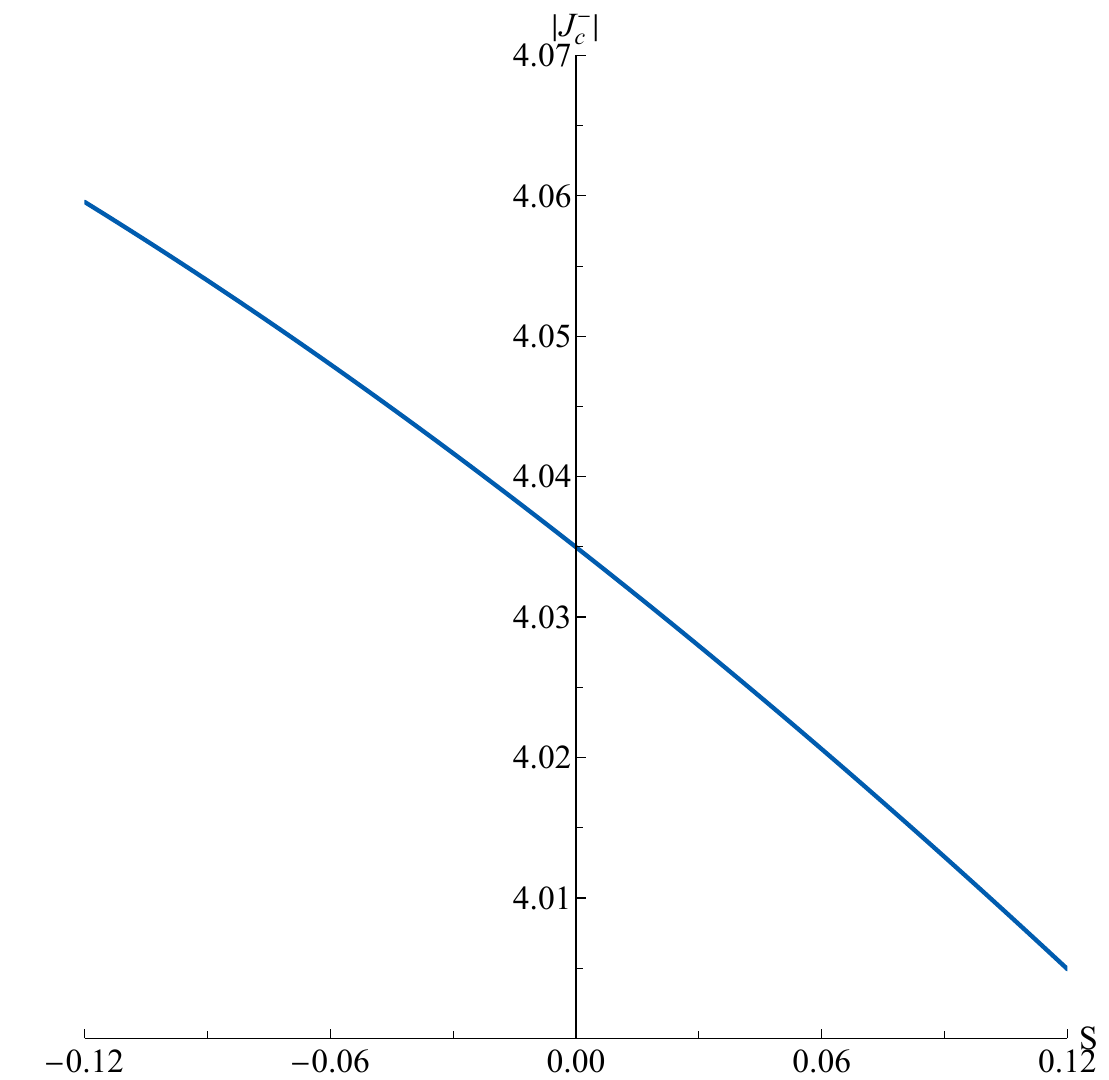}}\\[5mm]
	\subcaptionbox{}{\includegraphics[width=0.42\textwidth]{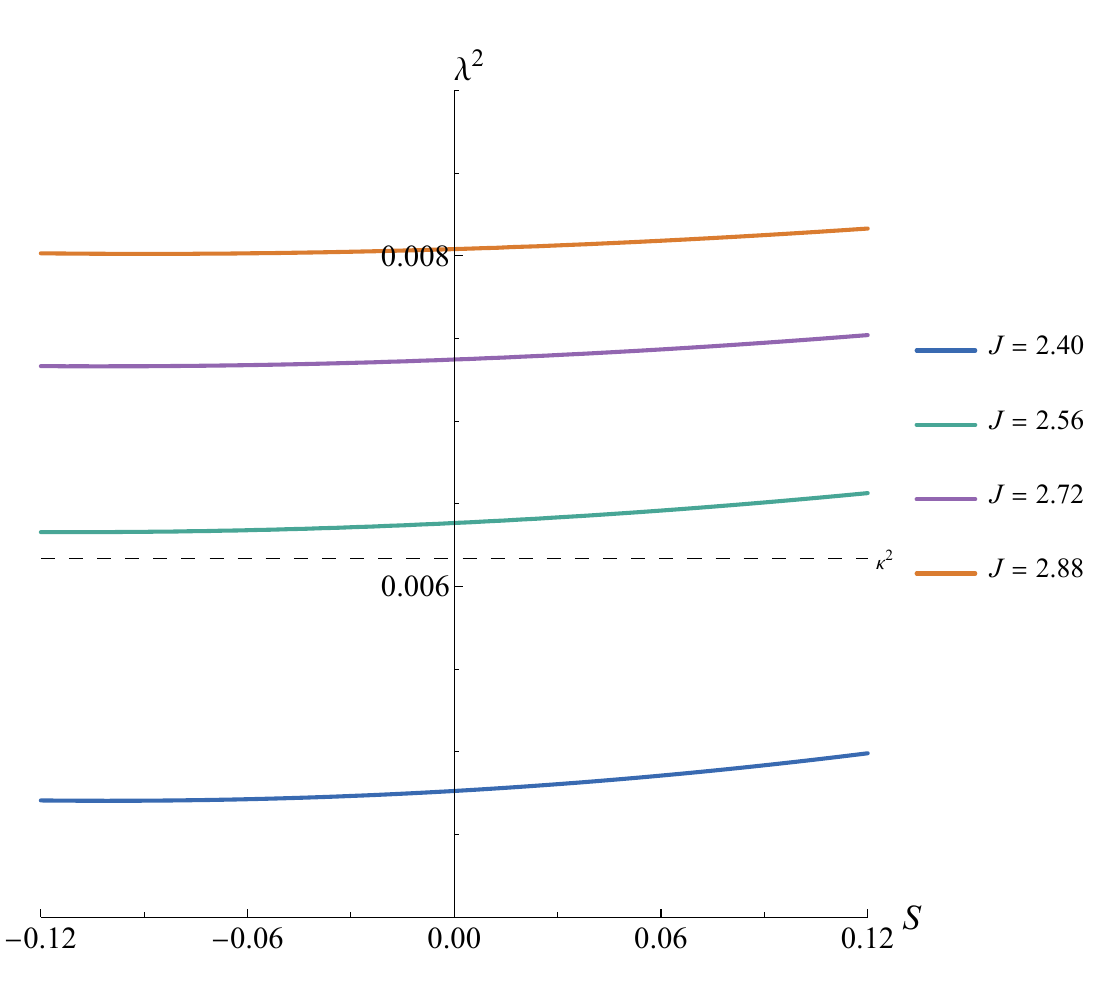}}
	\subcaptionbox{}{\includegraphics[width=0.42\textwidth]{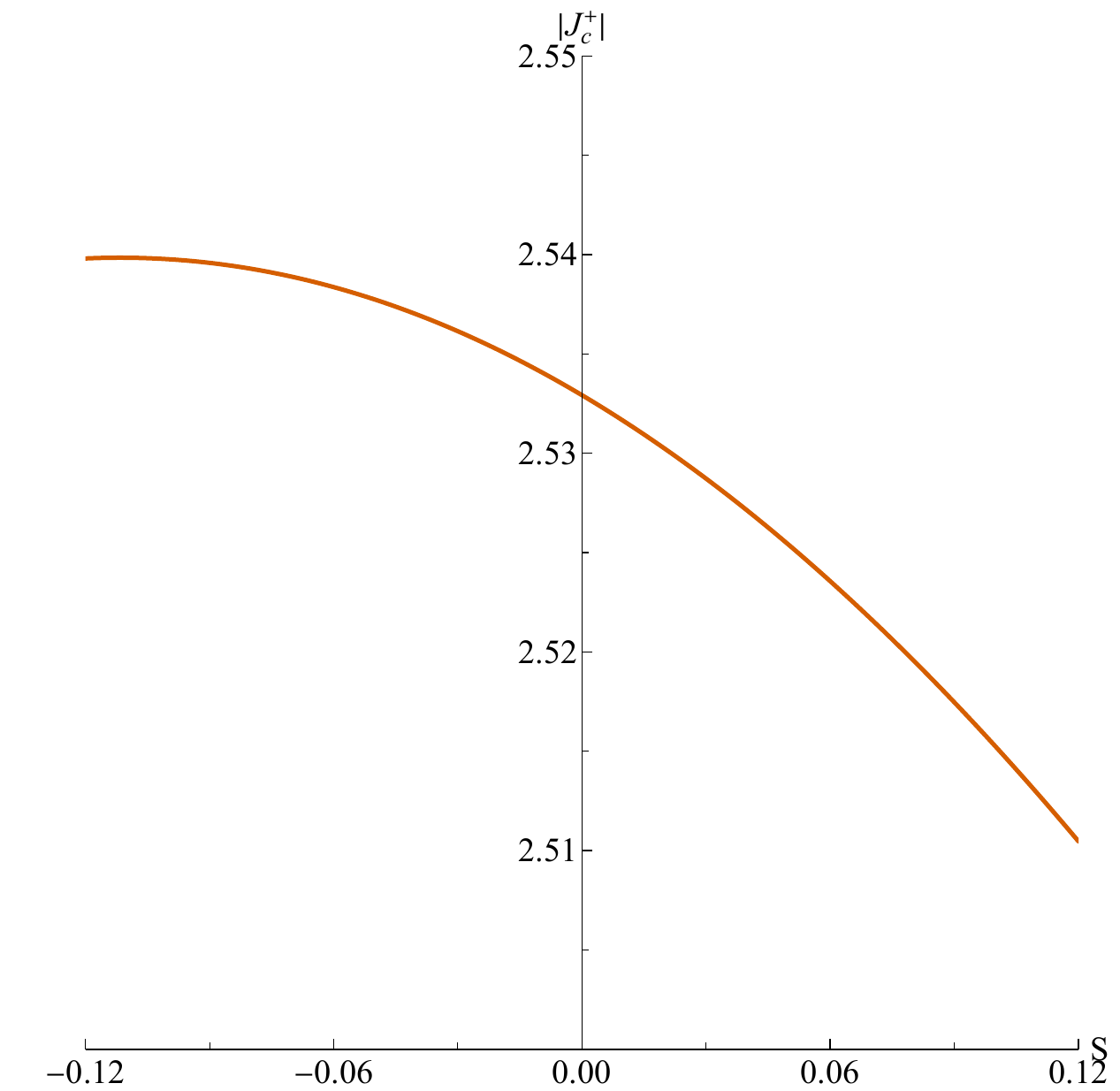}}
	\caption{Dependence of the squared LE and critical angular momentum on the particle spin for $\eta=0.90$.}
	\label{fig4}
\end{figure*}

\begin{figure*}[htbp]
	\centering
	\subcaptionbox{}{\includegraphics[width=0.42\textwidth]{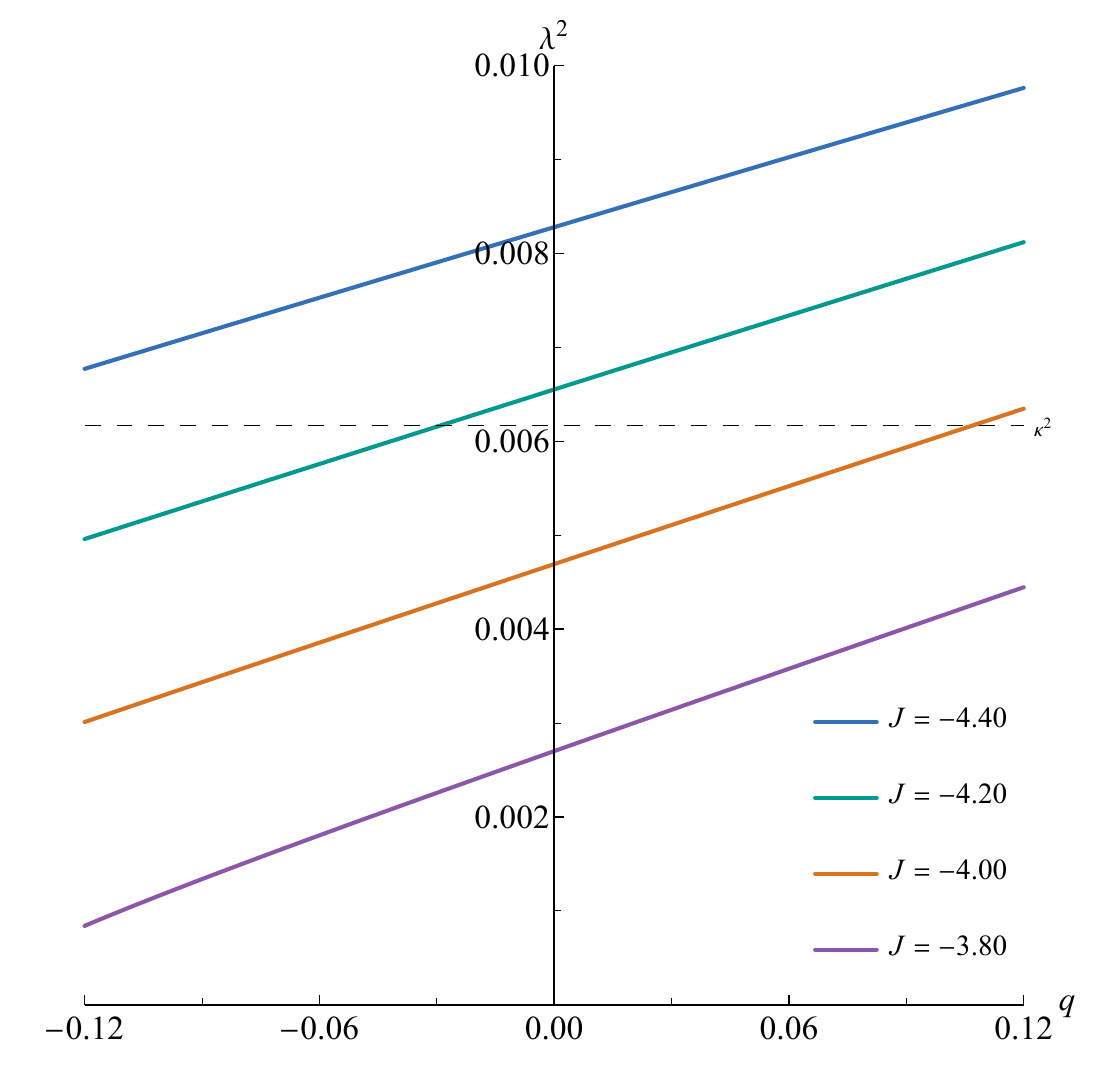}}
	\subcaptionbox{}{\includegraphics[width=0.42\textwidth]{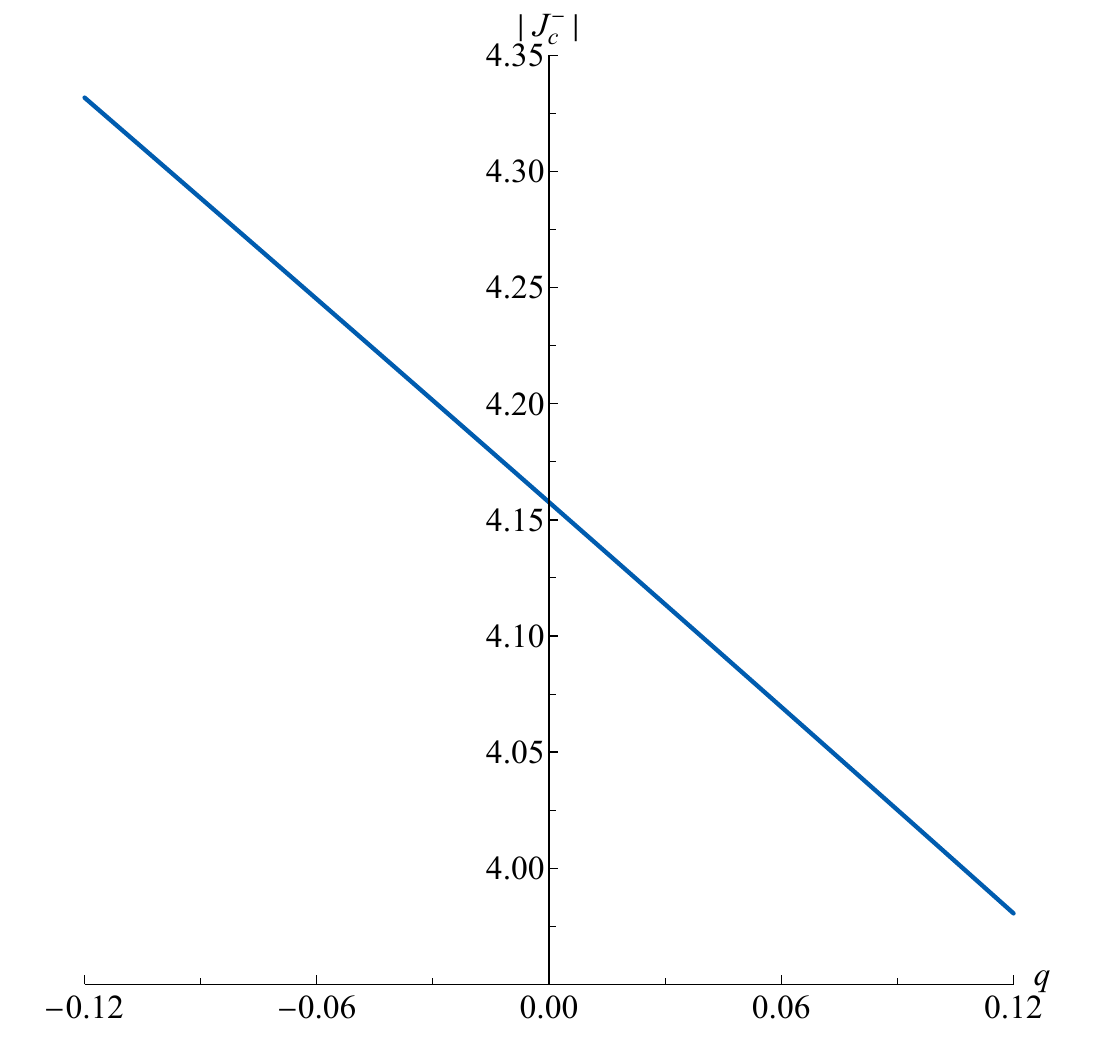}}\\[5mm]
	\subcaptionbox{}{\includegraphics[width=0.42\textwidth]{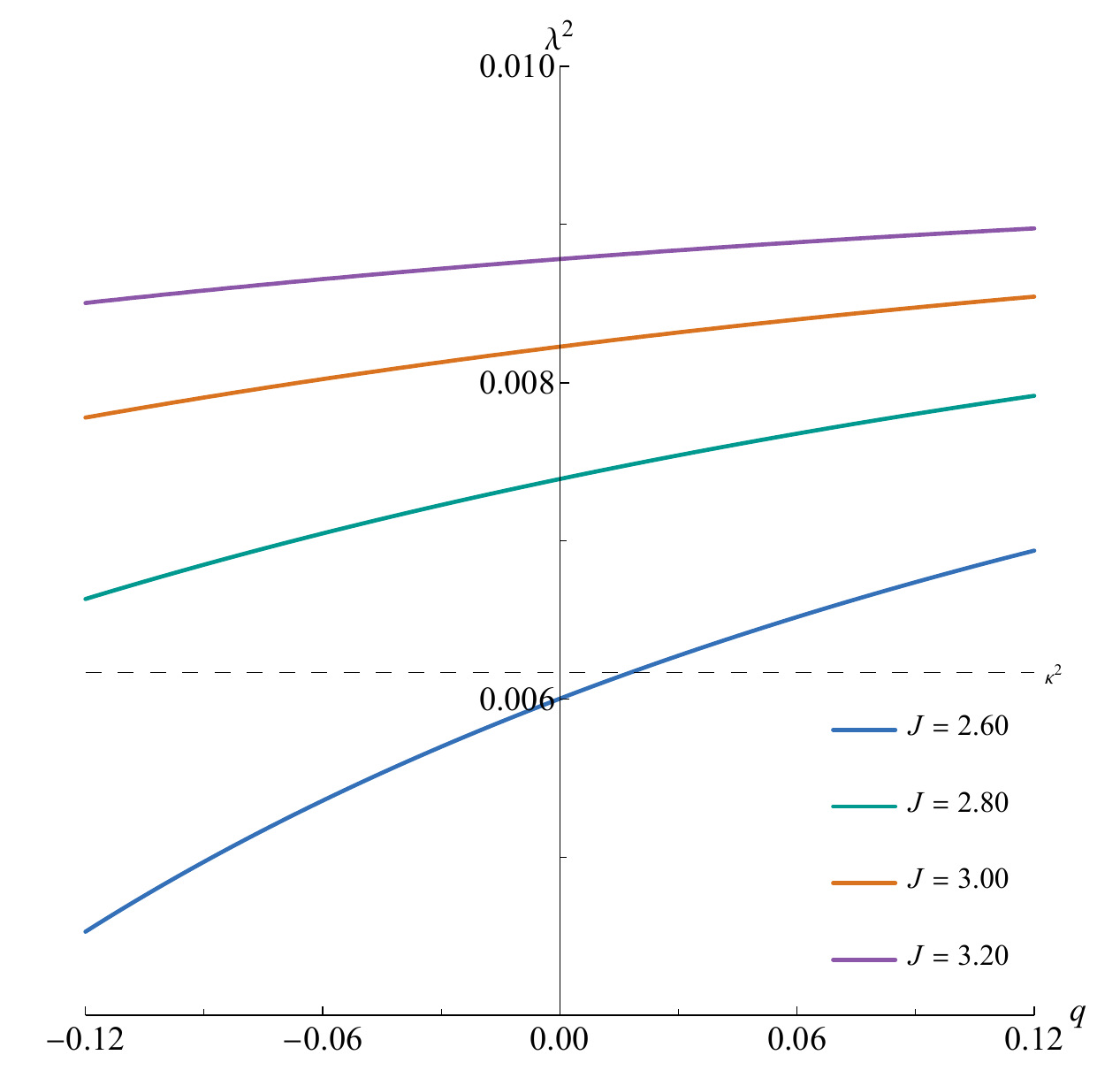}}
	\subcaptionbox{}{\includegraphics[width=0.42\textwidth]{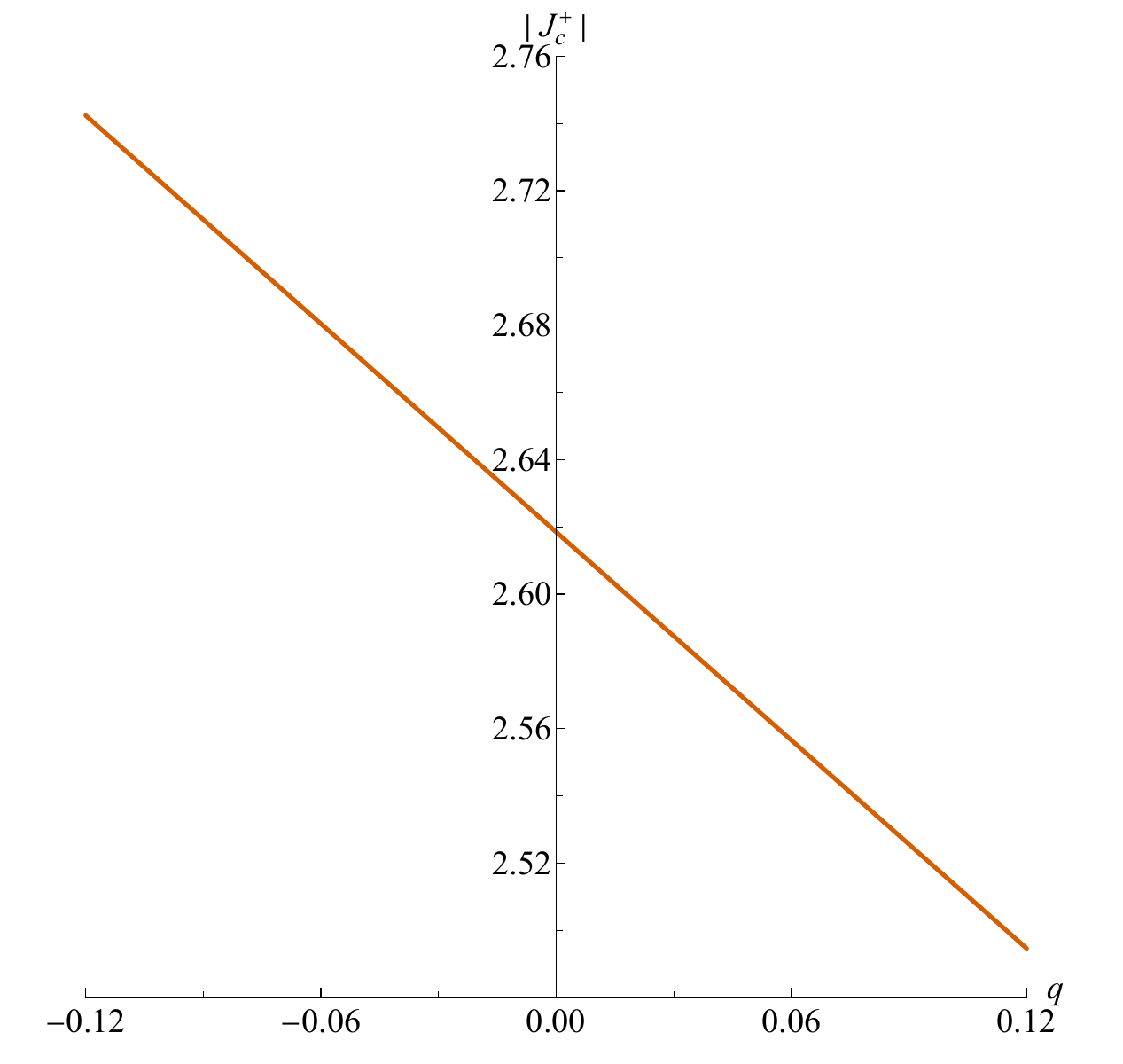}}
	\caption{Dependence of the squared LE and critical angular momentum on the particle charge for $\eta=0.90$.}
	\label{fig5}
\end{figure*}

Using Eqs.~\eqref{sf} and \eqref{le}, we numerically compare the LE characterizing the local radial instability with the black hole surface gravity and examine the corresponding chaos bound behavior. The numerical results are presented in Figs.~\ref{fig1}--\ref{fig5}. We define $\Delta_{\lambda\kappa}\equiv\lambda^{2}-\kappa^{2}$, such that $\Delta_{\lambda\kappa}\leq0$ corresponds to satisfaction of the bound, whereas $\Delta_{\lambda\kappa}>0$ signals violation. Unless stated otherwise, the parameters are fixed at $M=1.00$, $Q=0.77$, $a=0.50$, $S=0.10$, and $q=0.10$.

Figure~\ref{fig1} illustrates the effects of the dimensionless deformation parameter \(\eta\) on the orbital-instability scale, the surface gravity, and the coordinate separation between the unstable circular orbit and the outer horizon. Panels~(a) and (b) show the squared LE and the squared surface gravity as functions of the deformation parameter for $J=-6.00$ and $J=6.00$, respectively. In both cases, $\lambda^{2}$ and $\kappa^{2}$ decrease as the black-bounce regularization becomes stronger, with the latter decreasing substantially more rapidly. In the weakly deformed regime, $\kappa^{2}>\lambda^{2}$ and the chaos bound is therefore satisfied. As the regularization strength increases, the two quantities become equal at critical values $\eta=\eta_{01}$ and $\eta=\eta_{02}$ in panels (a) and (b), respectively, beyond which $\lambda^{2}>\kappa^{2}$ and the bound is violated. An important feature of this behavior is that the LE itself is not enhanced by the deformation; rather, it decreases together with the surface gravity. The onset of violation therefore does not result from an absolute enhancement of the local orbital-instability, but from the different responses of this instability and surface gravity scales to the black-bounce regularization. As the deformation increases, $\kappa^{2}$ decreases more rapidly than \(\lambda^2\), so that their ordering is eventually reversed. Equivalently, the surface gravity scale is suppressed more strongly than the local orbital-instability scale. Consequently, although the absolute growth rate of radial perturbations decreases, it becomes progressively larger relative to the surface gravity scale, eventually yielding $\lambda^{2}>\kappa^{2}$. Panels~(a) and (b) thus provide a clear example of a background-controlled route to chaos bound violation, in which the deformation alters the relative balance between the two characteristic scales. This behavior is consistent with the geometric mechanism discussed in the recent study \cite{TBZ} and shows how such a scale imbalance can arise in the black-bounce--Kerr--Newman geometry.

Panels~(c) and (d) show the coordinate separation $r_{0}-r_{+}$ between the unstable circular orbit and the outer horizon for the two angular momentum branches. In both cases, this separation increases monotonically with the deformation parameter. The increase is relatively mild in the weakly deformed regime but becomes progressively more pronounced as the regularization strengthens and the geometry approaches the black hole--wormhole transition. Thus, in the adopted radial coordinate, the unstable orbit does not move progressively closer to the horizon as the deformation increases; instead, the orbit--horizon separation grows. A pronounced branch dependence is also evident in its magnitude. Over the parameter range considered, the minimum value of $r_{0}-r_{+}$ is approximately $2.162$ for $J=-6.00$, whereas it is only about $0.264$ for $J=6.00$. The unstable orbit on the negative angular momentum branch therefore remains substantially farther from the outer horizon than that on the positive angular momentum branch. Despite this marked difference, both branches exhibit the same qualitative transition from the bound-satisfying to the bound-violating regime as the regularization increases. This comparison indicates that the onset of violation is not determined solely by the coordinate proximity of the unstable orbit to the horizon.

The four panels consistently point to a common physical picture governed by the competition between the orbital-instability and surface gravity scales. The black-bounce deformation reduces both the LE and the surface gravity, but suppresses the latter more strongly. Meanwhile, the increasing orbit--horizon separation shows that the bound-violating regime is not associated with a progressive approach of the unstable orbit to the horizon. The violation is therefore more naturally attributed to the deformation-induced imbalance between the two characteristic scales, rather than to an absolute enhancement of the local orbital instability or simply to the coordinate proximity of the unstable orbit to the horizon.

Figure~\ref{fig2} examines the influence of the black hole charge on chaos bound violation. Since the critical regularization scale \(\ell_+\) depends on the black hole charge, fixing \(\eta=\ell/\ell_+\) while varying the black hole charge would simultaneously change the physical regularization scale \(\ell\). We therefore fix several values of \(\ell\) and plot \(\Delta_{\lambda\kappa}\) as a function of the charge.
As shown in panel~(a), the deviation increases overall with the black hole charge along this angular momentum branch and changes sign once a charge-dependent threshold is crossed. Increasing the charge therefore drives the system progressively toward, and eventually into, the bound-violating regime. At fixed charge, a larger regularization scale generally produces a larger deviation, indicating that both the black hole charge and the black-bounce regularization tend to increase \(\Delta_{\lambda\kappa}\) along this branch. Since both parameters modify the horizon structure as well as the local geometry governing the unstable circular orbit, their combined effect changes the relative balance between the surface gravity and orbital-instability scales in a direction that favors violation. Panel~(a) therefore provides another example of a background-controlled realization of chaos bound violation, where variations of the background parameters modify the two characteristic scales and shift their relative balance toward the violating regime. At larger charge, the curves corresponding to different regularization scales gradually converge, indicating that the sensitivity of the bound behavior to the black-bounce regularization becomes progressively weaker over the parameter range considered. Panel~(b) shows the corresponding behavior after reversing the direction of the particle angular momentum while keeping its magnitude fixed. In contrast to the monotonic trend in panel~(a), the deviation first increases and then decreases as the charge grows. This nonmonotonic behavior indicates that the relative balance between the surface gravity and orbital-instability scales depends nontrivially on the angular momentum branch. At relatively small charge, their combined effect shifts the system toward larger deviations and hence toward the bound-violating regime. As the charge increases further, however, this tendency weakens and eventually reverses, leading to a decrease in $\Delta_{\lambda\kappa}$. The curves again approach one another in the large charge regime, showing that the influence of the regularization scale becomes progressively less pronounced.

Taken together, panels~(a) and (b) demonstrate that the effect of the black hole charge on the chaos bound is strongly branch dependent. On one angular momentum branch, increasing the charge and the regularization both shift the system toward violation, giving rise to an overall monotonic increase in $\Delta_{\lambda\kappa}$, whereas on the opposite branch their combined influence produces a nonmonotonic response. Despite this difference, both panels show a reduced sensitivity to the regularization scale at sufficiently large charge. Within the parameter range considered, the convergence of the curves indicates that the sensitivity of \(\Delta_{\lambda\kappa}\) to the regularization scale becomes progressively weaker as the black hole charge increases.

Figure~\ref{fig3} shows the deviation \(\Delta_{\lambda\kappa}=\lambda^2-\kappa^2\) as a function of the particle total angular momentum $J$. Since each curve corresponds to a fixed background, the surface gravity $\kappa$ remains unchanged. Therefore, the dependence of $\Delta_{\lambda\kappa}$ on the total angular momentum directly reflects the variation of squared LE. The behavior of the chaos bound along a given angular momentum scan is consequently determined by changes in the unstable circular orbit and its local radial instability, rather than by any variation of the surface gravity scale. Below a branch-dependent threshold in $|J|$, no admissible unstable circular orbit exists. Once this threshold is reached, a local maximum of the radial effective potential appears and an unstable circular orbit becomes available. Near the onset of each branch, the LE remains below the surface gravity and the chaos bound is satisfied. As $|J|$ increases, the local radial instability generally strengthens and $\lambda^2$ increases, eventually approaching or exceeding the squared surface gravity scale on those branches for which violation occurs. The increase is relatively pronounced at moderate angular momentum but becomes progressively weaker in the large-$|J|$ regime. These results show that, for a fixed spacetime background, the particle total angular momentum controls the local orbital instability by shifting the unstable circular orbit and modifying the local curvature of the effective potential. Since the surface gravity remains unchanged, the transition from the bound-satisfying to the bound-violating regime along a given angular momentum scan originates solely from the change in the instability scale. This behavior therefore identifies a probe-dynamics-controlled realization of chaos bound violation, in which the bound is crossed through the enhancement of the orbital-instability scale at fixed surface gravity.

The black-bounce regularization nevertheless has an important influence on this transition. Comparing curves corresponding to different values of the deformation parameter shows that, at a given angular momentum, \(\Delta_{\lambda\kappa}\) depends sensitively on the background geometry, while the critical angular momentum for chaos-bound violation also shifts as the regularization is varied. This reflects the fact that the deformation changes not only the location and local properties of the unstable circular orbit but also, when different backgrounds are compared, the horizon structure and the surface gravity. The two effects should therefore be distinguished. Along a fixed background scan, varying the angular momentum changes the local instability scale while leaving the surface gravity scale unchanged; changing the deformation parameter, by contrast, changes both the orbital-instability and surface gravity scales associated with the background and therefore shifts the angular momentum threshold for violation. Figure~\ref{fig3} thus provides a complementary view of the same competition identified in Figs.~\ref{fig1} and \ref{fig2}: the relative magnitude of the orbital-instability and surface gravity scales is determined jointly by the particle dynamics and the underlying spacetime geometry.

A pronounced directional dependence is also observed between the two angular momentum branches. As $|J|$ increases, admissible unstable circular orbits appear at different angular momentum magnitudes on the two branches, demonstrating that the orbit-existence condition itself depends on the direction of the particle motion relative to the rotating background. Within the range where both branches coexist, the branch directed opposite to the positive \(z\)-axis exhibits a smaller \(\lambda^2\) at relatively small \(|J|\), whereas this ordering reverses as \(|J|\) increases. This crossover reflects the combined dependence of the local instability on the angular momentum, the orbital location, and the curvature of the effective potential in the rotating geometry, rather than on the sign of angular momentum alone. Consequently, the two branches approach the surface gravity scale differently and exhibit distinct critical angular momenta for chaos bound violation. For example, at $\eta=0.00$, $0.25$, and $0.50$, the branch directed along the positive $z$-axis remains within the bound over the corresponding parameter range, whereas the opposite branch can already enter the bound-violating regime. The resulting asymmetry highlights the directional dependence of the particle dynamics in the rotating spacetime and explains why the two orbital branches respond differently to otherwise identical background parameters.

Figure~\ref{fig4} further examines the influence of the particle spin on the chaos bound behavior of the unstable circular orbits considered in Fig.~\ref{fig3}. Since the black hole background is held fixed, the surface gravity remains unchanged, and the spin dependence is entirely associated with modifications of the local orbital dynamics. As shown in panel~(a), the squared LE increases as the particle spin varies from negative to positive values along the first angular momentum branch. Within the spin convention adopted here, this behavior indicates that spin--curvature coupling enhances the radial instability for positive spin and suppresses it for negative spin. The instability also increases with the magnitude of the total angular momentum, and once a spin-dependent threshold is reached, the squared LE exceeds the squared surface gravity and the bound is violated. The effect of spin on the threshold is made explicit in panel~(b), where the critical angular momentum $|J^{-}_{\mathrm c}|$ decreases monotonically as the spin becomes more positive. Within the spin range considered and under the convention adopted here, positive spin enhances the local radial instability, whereas negative spin suppresses it. Panels~(c) and (d) display the corresponding behavior on the opposite angular momentum branch. The same qualitative trend persists: increasing the spin strengthens the local radial instability and lowers the critical angular momentum for violation. The quantitative response, however, differs between the two branches. This asymmetry reflects the directional dependence of the spin-induced corrections to the orbital dynamics in the rotating spacetime. Through spin--curvature coupling, variations in the spin modify the location and local properties of the unstable circular orbit, while the rotation of the background makes these corrections sensitive to the direction of the orbital motion. Consequently, the same change in spin produces different shifts in the exponent and in the critical angular momentum on the two branches.

Figure~\ref{fig4} thus shows that the particle spin acts as a systematic regulator of the local orbital instability at fixed background geometry, thereby shifting the critical angular momentum for the violation. Within the parameter range considered, the total angular momentum provides the main variation of the instability along the scans shown, whereas spin--curvature coupling regulates the local instability and shifts the angular momentum threshold at which the chaos bound is violated. Positive spin shifts this transition toward smaller angular momentum magnitudes, whereas negative spin shifts it in the opposite direction. The magnitude of this spin-induced shift is itself branch dependent, further indicating that the response of the unstable orbit is sensitive to the orientation of the particle motion and spin relative to the background rotation. Thus, the spin does not constitute a separate mechanism from the probe-dynamics-controlled route; rather, through spin--curvature coupling, it acts as a branch-dependent regulator of the local orbital-instability scale and thereby modifies how readily this scale reaches and exceeds the fixed surface gravity scale.

Figure~\ref{fig5} examines the influence of the particle charge on the chaos bound behavior of unstable circular orbits for a fixed black hole background. Since the horizon geometry and surface gravity remain unchanged, the charge dependence arises entirely from the modification of the orbital dynamics through the electromagnetic coupling. Panel~(a) corresponds to the branch for which the particle total angular momentum is directed opposite to the positive $z$-axis. For all angular momenta considered, the squared LE increases monotonically as the particle charge varies from negative to positive values. Within the charge convention and parameter range adopted here, a more positive particle charge therefore strengthens the local radial instability on this branch. For $J=-3.80$, the exponent remains below the surface gravity throughout the charge interval, whereas for $J=-4.40$ the bound is violated over the entire interval. At intermediate angular momenta, the two scales cross at a charge-dependent threshold, showing that variation of the particle charge alone can drive the unstable orbit from the bound-satisfying to the bound-violating regime while the surface gravity remains fixed. This trend is quantified in panel~(b), where the critical angular momentum $|J^{-}_{\mathrm c}|$ decreases as the particle charge becomes more positive. The electromagnetic interaction therefore lowers the angular momentum threshold for violation in the positive charge direction, whereas more negative charge shifts the transition toward larger $|J|$.

Panels~(c) and (d) show the corresponding behavior on the opposite angular momentum branch. The same qualitative charge dependence persists: increasing the particle charge enhances the local radial instability and reduces the critical angular momentum at which the bound is crossed. The quantitative response, however, differs between the two branches. The quantitative sensitivity of squared LE to the particle charge differs between the two angular momentum branches. This asymmetry indicates that the charge-induced modification of the unstable orbit dynamics is sensitive to the direction of the particle motion relative to the rotating background. The resulting instability is therefore governed by the combined dependence on the particle charge and angular momentum, together with the branch-dependent orbital location and local structure of the effective potential in the rotating geometry.

Figure~\ref{fig5} therefore demonstrates that, at fixed background geometry, the electromagnetic interaction associated with the particle charge modifies the local orbital instability and consequently shifts the threshold for chaos bound violation. A more positive charge facilitates the transition by increasing the local instability and lowering the critical $|J|$, whereas a more negative charge delays its onset. The magnitude of this charge-induced shift is branch dependent, further illustrating the directional sensitivity of charged-particle dynamics in a rotating spacetime. Thus, as in the case of particle spin, the charge does not constitute a separate mechanism from the probe-dynamics-controlled route; rather, it acts as a branch-dependent electromagnetic regulator of the local orbital-instability scale and thereby modifies how readily this scale reaches and exceeds the fixed surface-gravity scale. Taken together, these results show that, within the parameter ranges considered, the total angular momentum produces the dominant variation in the local radial instability, while the particle spin and charge mainly shift the instability and the corresponding threshold for chaos bound violation.

\section{Discussion and conclusion}
\label{sec4}

In this paper, we have investigated the local radial instability of charged spinning test particles on unstable circular orbits in the black-bounce--Kerr--Newman spacetime and examined the chaos bound by comparing the LE with the black hole surface gravity. Our results indicate that the behavior of the bound is controlled by the competition between the orbital-instability and  horizon scales. Depending on whether the background or particle parameters are varied, the system can cross the threshold \(\lambda^{2}=\kappa^{2}\) through either a background-controlled or a probe-dynamics-controlled route.

When the black hole background is varied, both the surface gravity scale and the unstable orbit dynamics respond to the change in the spacetime geometry. An important result is that chaos bound violation does not necessarily require an enhancement of the local orbital instability. In particular, increasing the black-bounce regularization reduces both the squared LE and squared surface gravity, with the squared surface gravity decreasing more rapidly than the squared LE. Consequently, the system can cross from $\lambda^{2}<\kappa^{2}$ to $\lambda^{2}>\kappa^{2}$ even though the unstable orbit itself becomes locally less unstable. The violation is therefore governed by the relative response of the two characteristic scales rather than by the absolute magnitude of the LE alone. Moreover, the increase in the coordinate separation \(r_0-r_+\) shows that this behavior cannot be attributed simply to a progressive approach of the unstable orbit toward the event horizon. This behavior is conceptually related to the geometric mechanism emphasized in Ref.~[32], while the present analysis demonstrates how an analogous imbalance between the orbital and horizon scales arises in the black-bounce--Kerr--Newman geometry. 

A complementary behavior emerges when the black hole background is held fixed. In this case, the horizon geometry and hence the surface gravity remain unchanged, so that the variation of $\lambda^{2}-\kappa^{2}$ is entirely controlled by the particle dynamics. Within the parameter ranges considered, the total angular momentum produces the dominant variation in the local radial instability and can drive the system across the threshold $\lambda^{2}=\kappa^{2}$. The particle spin and charge further modify the local orbital instability through spin--curvature and electromagnetic couplings, respectively, thereby shifting both the LE and the critical conditions for the violation. These effects are branch dependent because the response of the unstable orbit dynamics to spin and charge depends on the direction of the particle motion relative to the rotating background, leading to different instability strengths and violation thresholds for the two orbital branches. Thus, at fixed background, the bound can be violated through a probe-dynamics-controlled route while the horizon structure and surface gravity remain unchanged.

These two behaviors are different manifestations of the same underlying two-scale competition. Background deformations can move both the surface gravity and LE and produce a crossing when their responses become sufficiently unequal, whereas variations of the particle degrees of freedom primarily modify the orbital-instability scale at fixed surface gravity. Hence, the condition $\lambda^{2}>\kappa^{2}$ should not in general be interpreted as evidence for an absolute enhancement of the local orbital instability. It may instead result either from a stronger suppression of the surface gravity scale or from an enhancement of the orbital-instability scale relative to a fixed horizon scale. This unified picture provides a common interpretation of the different parameter dependences found in the background and probe sectors.
Finally, the condition $\lambda^{2}>\kappa^{2}$ characterizes only the local radial instability of an unstable circular orbit within the symmetry-reduced sector considered here and does not by itself establish global chaotic dynamics in the full phase space. Our analysis is restricted to equatorial motion with a prescribed spin configuration and therefore does not include generic polar motion, spin precession, or couplings among radial, vertical, and spin perturbations. Extending the analysis to generic nonequatorial trajectories and to coupled perturbations of the full MPD system would provide a more complete test of whether the local instability behavior identified here persists in the global dynamics of charged spinning particles.


\begin{thebibliography}{99}
	
	\small
	
\bibitem{HT}
K. Hashimoto and N. Tanahashi, 
\emph{Universality in chaos of particle motion near black hole horizon}, 
\emph{Phys. Rev.} \textbf{D 95} (2017) 024007.

\bibitem{MSS}
J. Maldacena, S.H. Shenker and D. Stanford, 
\emph{A bound on chaos}, 
\emph{JHEP} \textbf{1608} (2016) 106. 

\bibitem{SS}
S.H. Shenker and D. Stanford,
\emph{Black holes and the butterfly effect},
\emph{JHEP} \textbf{1403} (2014) 067.

\bibitem{RS}
D.A. Roberts and D. Stanford,
\emph{Diagnosing chaos using four-point functions in two-dimensional conformal field theory},
\emph{Phys. Rev. Lett.} \textbf{115} (2015) 131603.

\bibitem{MS}
J. Maldacena and D. Stanford,
\emph{Remarks on the Sachdev--Ye--Kitaev model},
\emph{Phys. Rev.} \textbf{D 94} (2016) 106002.

\bibitem{SSM}
S. H. Shenker and D. Stanford,
\emph{Multiple shocks},
\emph{JHEP} \textbf{1412} (2014) 046.

\bibitem{RSS}
D. A. Roberts, D. Stanford, and L. Susskind,
\emph{Localized shocks},
\emph{JHEP} \textbf{1503} (2015) 051.

\bibitem{SSS}
S. H. Shenker and D. Stanford,
\emph{Stringy effects in scrambling},
\emph{JHEP} \textbf{1505} (2015) 132.

\bibitem{ZLL}
Q.Q. Zhao, Y.Z. Li and H. L\"u, 
\emph{Static equilibria of charged particles around charged black holes: Chaos bound and its violations}, 
\emph{Phys. Rev.} \textbf{D 98} (2018) 124001.

\bibitem{LG1}
Y.Q. Lei and X.H. Ge, 
\emph{Circular motion of charged particles near charged black hole}, 
\emph{Phys. Rev.} \textbf{D 105} (2022) 084011.

\bibitem{LG2}
Y.Q. Lei, X.H. Ge and C. Ran, 
\emph{Chaos of particle motion near a black hole with quasitopological electromagnetism},
\emph{Phys. Rev.} \textbf{D 104} (2021) 046020.

\bibitem{KG1}
N. Kan and B. Gwak, 
\emph{Bound of Lyapunov exponent in Kerr-Newman black holes via charged particle}, 
\emph{Phys. Rev.} \textbf{D 105} 026006 (2022).

\bibitem{KG2}
B. Gwak, N. Kan, B.H. Lee and H. Lee, 
\emph{Violation of bound on chaos for charged probe in Kerr-Newman-AdS black hole}, 
\emph{JHEP} \textbf{2209} (2022) 026.

\bibitem{KG3}
H. Lee and B. Gwak, 
\emph{Bound on Lyapunov exponent for a charged particle in Kerr-Sen-AdS black hole}, 
\emph{Phys. Rev.} \textbf{D 112} (2025) 046018.

\bibitem{KG5}
H. Lee and B. Gwak, 
\emph{Frame dependence of bound on Lyapunov exponent in Dilatonic Reissner-Nordström-AdS and Kerr-Sen-AdS black holes}, 
arXiv:2510.16479 [gr-qc]

\bibitem{KG4}
J. Park and B. Gwak, 
\emph{Bound on Lyapunov exponent in Kerr-Newman-de Sitter black holes by a charged particle},
\emph{JHEP} \textbf{2404} (2024) 023.

\bibitem{GCYW2}
D.Y. Chen and C.H. Gao, 
\emph{Angular momentum and chaos bound of charged particles around Einstein-Euler-Heisenberg AdS black holes}, 
\emph{New J. Phys.} \textbf{24} (2022) 123014.


\bibitem{FAAC1}
A. Fayyaz, G. Abbas, M.B. Asfour and D.Y. Chen,  
\emph{Chaotic dynamics and bound violation in 4D EGB-AdS black holes with massive gravitons}, 
\emph{Phys. Lett.} \textbf{B 871} (2025) 139965.

\bibitem{FAAC2}
Y. Xie, J. Wang and B. Tang, 
\emph{Circular motion and chaos bound of a charged particle near charged 4D Einstein–Gauss–Bonnet-AdS black holes}, 
\emph{Phys. Dark. Univ} \textbf{42} (2023) 101271.

\bibitem{YCL}
C. Yang, D.Y. Chen and Y. Liu, 
\emph{Motions of spinning particles and chaos bound in Reissner-Nordstr\"om spacetime}, 
\emph{JHEP} \textbf{04} (2026) 205.

\bibitem{LCL1}
X. Li, B.B. Chen and G.P. Li, 
\emph{Testing the chaos bound in the spinor field of Einstein–Euler–Heisenberg–Anti-de Sitter spacetime}, 
arXiv:2604.03914[gr-qc].

\bibitem{LG4}
Y.Q. Lei and X.H. Ge, 
\emph{Stationary equilibrium of test particles near charged black branes with the hyperscaling violating factor}, 
\emph{Phys. Rev.} \textbf{D 107} (2023) 106002.

\bibitem{DPS}
P. Dutta, K.L. Panigrahi and B. Singh, 
\emph{Chaos bound and its violation in black p-brane}, 
\emph{JHEP} \textbf{02} (2025) 043.

\bibitem{JLLL}
S. Jeong, B.H. Lee, H. Lee and W. Lee, 
\emph{Homoclinic orbit and the violation of the chaos bound around a black hole with anisotropic matter fields}, 
\emph{Phys. Rev.} \textbf{D 107} (2023) 104037.

\bibitem{LTW1}
F.H. Lü, J. Tao and P. Wang, 
\emph{Minimal length effects on chaotic motion of particles around black hole horizon}, 
\emph{JCAP} \textbf{1812} (2018) 036.

\bibitem{LTW2}
X.B. Guo, K.K. Liang, B.R. Mu, P. Wang and H.T. Yang, 
\emph{Minimal length effects on motion of a particle in Rindler space}, 
\emph{Chin. Phys.} \textbf{C} 45 (2021) 023115.

\bibitem{SPN}
B. Singh, N. Padhi and R.R. Nayak, 
\emph{Circular orbits and chaos bound in slow-rotating curved acoustic black holes}, 
\emph{Eur. Phys. J.} \textbf{C 85} (2025) 570.

\bibitem{DG}
D. Giataganas, 
\emph{Chaotic motion near black hole and cosmological horizons}, 
\emph{Fortsch. Phys.} \textbf{70} (2022) 2200001.

\bibitem{RP}
R. Pourkhodabakhshi, 
\emph{Saturation of chaos bound in a phenomenological rotating black hole--effective matter system at low-temperature limit}, 
\emph{Phys. Rev.} \textbf{D 112} (2025) 064085.

\bibitem{TBAZ}
T.V. Targema, K. Bamba, R. Ali and U. Zafar, 
\emph{Physical constraints on the MSS chaos-bound in black hole spacetimes}, 
\emph{Phys. Rev.} \textbf{D 113} (2026) 104056.

\bibitem{TBZ}
T.V. Targema, K. Bamba and U. Zafar, 
\emph{A universal geometric mechanism for chaos-bound violations in black hole spacetimes},
arXiv:2605.26829[hep-th].

\bibitem{HS}
K. Hashimoto and K. Sugiura, 
\emph{Causality bounds chaos in geodesic motion}, 
\emph{Phys. Rev.} \textbf{D 107} (2023) 066005.

\bibitem{TM}
T. Morita, 
\emph{Bound on Lyapunov exponent in c=1 matrix model}, 
\emph{Eur. Phys. J.} \textbf{C 80}  (2020)  331.

\bibitem{DMM}
S. Dalui, B.R. Majhi and P. Mishra, 
\emph{Presence of horizon makes particle motion chaotic}, 
\emph{Phys. Lett.} \textbf{B 788} (2019) 486.

\bibitem{HMTW}
K. Hashimoto, K. Murata, N. Tanahashi and R. Watanabe, 
\emph{A bound on energy dependence of chaos}, 
\emph{Phys. Rev.} \textbf{D  106} (2022)  126010.

\bibitem{GT}
E. Gallo and T. Mädler, 
\emph{Bounds for Lyapunov exponent of circular light orbits in black holes}, 
\emph{Eur. Phys. J.} \textbf{C 85} (2025) 299.

\bibitem{LG3}
Y.Q. Lei, X.H. Ge  and S. Dalui, 
\emph{Thermodynamic stability versus chaos bound violation in D-dimensional RN black holes: Angular momentum effects and phase transitions}, 
\emph{Phys. Lett.} \textbf{B 856} (2024) 138929.

\bibitem{GCYW1}
C.H. Gao, D.Y. Chen, C.Y. Yu and P. Wang, 
\emph{Chaos bound and its violation in charged Kiselev black hole}, 
\emph{Phys. Lett.} \textbf{B 833} (2022) 137343.

\bibitem{LMX}
K. Li, D.Z. Ma and Z.M. Xu, 
\emph{Chaotic dynamics of string around the charged Kiselev black hole}, 
\emph{Phys. Lett.} \textbf{B 860} (2025) 139164.

\bibitem{HH}
R. Hojman and S. Hojman, 
\emph{Spinning charged test particles in a Kerr–Newman background}, 
\emph{Phys. Rev.} \textbf{D 15} (1977) 2724.

\bibitem{SV}
A. Simpson and M. Visser,
\emph{Black-bounce to traversable wormhole}, 
\emph{JCAP} \textbf{02} (2019) 042.

\bibitem{FLMSV}
E. Franzin, S. Liberati, J. Mazza, A. Simpson and M. Visser,
\emph{Charged black-bounce spacetimes}, 
\emph{JCAP} \textbf{07} (2021) 036. 

\bibitem{Ghosh}
S. Ghosh and A. Bhattacharyya,
\emph{Analytical study of gravitational lensing in Kerr-Newman black-bounce spacetime},
\emph{JCAP} \textbf{11} (2022) 006.

\bibitem{Li}
Q. Li, Q. Wang, and J. Jia,
\emph{On-axis absorption and scattering of charged massive scalar waves by Kerr-Newman black-bounce spacetime},
\emph{Phys. Rev.} \textbf{D 112} (2025) 124033.

\bibitem{WT}
W. Tulczyjew, 
\emph{Motion of multipole particles in general relativity theory}, 
\emph{Acta Physica Polonica} \textbf{18} (1959) 393.

\bibitem{HR}
A.J. Hanson and T. Regge, 
\emph{The relativistic spherical top}, 
\emph{Ann. Phys.} \textbf{87} (1974) 498. 

\bibitem{AO}
A. Ohashi, 
\emph{Multipole particle in relativity}, 
\emph{Phys. Rev.} \textbf{D 68} (2003) 044009.

\bibitem{KS}
K. Kyrian and O. Semerak, 
\emph{Spinning test particles in a Kerr field}, 
\emph{Mon. Not. R. Astron. Soc.} \textbf{382} (2007) 1922.

\bibitem{ZGWYL}
Y.P. Zhang, Y.B. Zeng, Y.Q. Wang, S.W. Wei and Y.X. Liu, 
\emph{Equatorial orbits of spinning test particle in rotating boson star}, 
\emph{Eur. Phys. J.} \textbf{C 82} (2022) 809.

\bibitem{ZW}
Y.K. Zhang and S.W. Wei, 
\emph{Effects of magnetic fields on spinning test particles orbiting Kerr-Bertotti-Robinson black holes}, 
\emph{Phys. Rev.} \textbf{D 113} (2026) 104024.

\bibitem{CMBWZ}
V. Cardoso, A.S. Miranda, E. Berti, H. Witek and V.T. Zanchin, 
\emph{Geodesic stability, Lyapunov exponents and quasinormal modes},  
\emph{Phys. Rev.} \textbf{D 79} (2009) 064016.

\bibitem{CHL}
S.Y. Ciou, T. Hsieh and D.S. Lee, 
\emph{Dynamics of spinning particles in Reissner-Nordstr\"om black hole exterior},  
\emph{JCAP} \textbf{05} (2025) 086.




\end{thebibliography}
\end{document}